\documentclass[a4paper,fleqn]{elsarticle}

\usepackage{amssymb}

\usepackage{graphicx}
\usepackage{dcolumn}
\usepackage{bm}
\usepackage{hyperref}

\usepackage[utf8]{inputenc}
\usepackage[T1]{fontenc}
\usepackage{geometry}
\usepackage{amsmath}
\usepackage{amssymb}
\usepackage{mathrsfs}
\usepackage{setspace}
\usepackage{mathalfa}
\usepackage{calrsfs}
\usepackage{tikz}
\usepackage{undertilde}
\usepackage{cancel}

\DeclareMathAlphabet\mathbfcal{OMS}{cmsy}{b}{n}
\journal{Annals of Physics}

\begin{document}

\begin{frontmatter}



\title{Hydrodynamic interpretation of generic squeezed coherent states: 
A kinetic theory}


\author[inst1]{Nezihe Uzun}

\affiliation[inst1]{organization={Univ Lyon, Ens de Lyon, Univ Lyon1, CNRS, Centre de Recherche Astrophysique de
Lyon, 46 allee d’Italie},
            city={Lyon},
            postcode={69007}, 
            state={},
            country={France}}

\begin{abstract}
The hydrodynamic interpretation of quantum mechanics treats a system of particles in an effective manner which allows one to study the system in a statistical fashion. In this work, we investigate squeezed coherent states within the hydrodynamic interpretation. The Hamiltonian operator in question is time dependent, $n$--dimensional and in quadratic order. We start by deriving a phase space Wigner probability distribution and an associated equilibrium entropy for the squeezed coherent states. Then, we decompose the joint phase space distribution into two portions: a marginal position distribution and a momentum distribution that is conditioned on the post--selection of positions. Our conditionally averaged momenta are shown to be equal to the Bohm's momenta whose connection to the weak measurements is already known. We also keep track of the corresponding classical system evolution by identifying shear, magnification and rotation components of the symplectic phase space dynamics. This allows us to pinpoint which portion of the underlying classical motion appears in which quantum statistical concept. We show that our probability distributions satisfy the Fokker--Planck equations exactly and they can be used to decompose the equilibrium entropy into the missing information in positions and in momenta as in the Sackur--Tetrode entropy of the classical kinetic theory. Eventually, we define a quantum pressure, a quantum temperature and a quantum internal energy which are related to each other in the same fashion as in the classical kinetic theory. We show that the quantum potential incorporates the kinetic part of the internal energy and the fluctuations around it. This allows us to suggest a quantum conditional virial relation. In the end, we show that the kinetic internal energy is linked to the fractional Fourier transformer part of the underlying classical dynamics similar to the case where the energy of a quantum oscillator is linked to its Maslov index.
\end{abstract}



\begin{keyword}
squeezed coherent states \sep hydrodynamic interpretation \sep quantum statistical mechanics \sep quantum phase space \sep Wigner function
\PACS 05.30.-d \sep 03.65.Sq
\end{keyword}

\end{frontmatter}


\section{Introduction}
The search for the underlying ties between classical and quantum theories has been a long quest without a unique solution. Nevertheless, for a given quantum system, certain states can be linked to its classical dynamics. For example, the wave function of a coherent state represents a system with minimum uncertainty in which the expectation values of the position and momentum operators follow classical trajectories. This is the closest one can get to a classical picture which was the original motivation of Schr{\"{o}}dinger when he first discovered those states \cite{Schrodinger:1926}. The name ``coherent'' was given by Glauber when he extended Schr{\"{o}}dinger's work for quantum optics in order to study the coherency properties of light  within the realm of standard harmonic oscillators \cite{Glauber:1963tx, Glauber:1963fi}. Since then, the definition of those states have been extended for more generic systems \cite{Perelomov:1986, Littlejohn:1985, Folland:1989}. They are specifically useful when studying the semiclassical limit of generic systems, or the ones which can be modelled by quadratic Hamiltonians \cite{Littlejohn:1985}.
Applications of squeezed coherent states are mostly known to the researchers within the quantum optics and quantum information processing fields. Those states are known to reduce the quantum mechanical noise and to increase the sensitivity in interferometric measurements \cite{Caves:1981,Walls:1983,Wu:1987,Ritze:1987}. This makes them perfect candidates for real life experiments that require highest level of precision. In addition, the extensive application area of squeezed coherent states provides links to even wider perspectives through their usage in gravitational wave detection \cite{Tse:2019,Acernese:2019,Lough:2021}, in bio--imaging \cite{Li:2020,Lawrie:2020} and even in the early universe cosmology \cite{Grishchuk:1990bj,Albrecht:1992kf,Martin:2012pea}. 

Alternatively, there exists a formalism in non--relativistic quantum mechanics that relates the wave function of a given system to the Hamilton--Jacobi equations of classical trajectories. This is known as causal/pilot wave/deterministic/the de Broglie--Bohm theory. In the literature, it is usually attributed to the work of Bohm \cite{Bohm:1952_I,Bohm:1952_II}, due to his major contribution and him reviving the pilot wave theory of de Broglie \cite{DeBroglie:1925,DeBroglie:1927} quite outspokenly. The early works of Madelung \cite{Madelung:1927}, in addition to the contributions of Takabayasi \cite{Takabayasi:1954} are usually overlooked.
Even though Madelung's, de Broglie's and Bohm's approaches are mathematically equivalent, their ontologies seem to be different.
According to the de Broglie--Bohm interpretation, particles follow trajectories guided by a pilot wave whose wave function has a physical connotation. This is true even for a single particle. However, in the hydrodynamic interpretation of Madelung, the idea is to consider many particles in which the corresponding wave function of the system reflects an effective dynamics. Throughout this work, we are going to favour Madelung's hydrodynamic interpretation as we would like to study the statistical behaviour of a given system. We will, nevertheless, refer to the underlying mathematical approach as  de Broglie--Madelung--Bohm theory. Note that in both of those interpretations, it is the so-called \textit{quantum potential} that is responsible for the underlying quantum phenomena which we will also study in detail.

Another course of action is the Wigner--Weyl--Moyal approach which allows one to define a quasi--probability distribution on a phase space analogous to the one in classical mechanics. For pure states, for example, Wigner function \cite{Wigner:1932} is shown to be the Weyl symbol of the density operator. This allows one to have a clear passage to classical physics as the Wigner function is used to obtain the expectation values of operators on phase space. Mainly, when combined with the Moyal product and the Moyal bracket \cite{Moyal:1949}, one recovers the von Neumann equation which reduces to the Liouville equation in the classical limit \cite{Hiley:2015xy}. This is at the core of the deformation quantization which provides connections between classical and quantum physics.

On the other hand, finding the exact solutions for time dependent systems both in classical and in quantum mechanics is not an easy task. As the invariants provide symmetries of a given system, the Lewis--Riesenfeld invariant method \cite{Lewis:1967,LewisRiesenfeld:1969} has been used widely in the literature in order to find the exact solutions of time dependent systems. The classical correspondent of this quantum operator is known to be an invariant of the associated classical system. Unfortunately, the Lewis--Riesenfeld invariant is usually considered as a mathematical tool. Its physical and/or geometric interpretation are not studied in detail in the literature \footnote{See \ref{appendix:Invariants} for some exceptions.}.

All of the approaches listed above provide means to relate classical and quantum mechanics. Naturally, the links between some of those formalisms have been already established. For example, one of the most concise investigations of the squeezed coherent states that makes use of the Wigner--Weyl--Moyal approach was given by Littlejohn \cite{Littlejohn:1985}. Moreover, the relationship between the Wigner--Weyl--Moyal approach and the de Broglie--Madelung--Bohm theory has been investigated before \cite{Hiley:2006,Hiley:2010,Hiley:2012w,Colomes:2015}. The Gaussian states of a harmonic oscillator \cite{Barut:1990,Dey:2013, Durr:2010} and a particle in 1-dimensional P{\"{o}}schl–Teller potential \cite{Dey:2013} have also been investigated within the de Broglie--Madelung--Bohm theory. In addition, the Lewis--Riesenfeld invariant of the Gaussian states was established in \cite{Yeh:1993} for one dimensional systems. However, there exists no study that unites all of the approaches mentioned above in order to present a full picture. 

In this work, we investigate time dependent squeezed coherent states in $n$--dimensions within a non--relativistic setting. It is known that a coherent state is an eigenstate of the annihilation operator. Thus, in general, there is a common practice to study them by making use of the ladder operators and the number states. We will not follow this route as our aim is to study a system within the hydrodynamic interpretation. We would like to keep track of the classical and the quantum phase space variables that are directly meaningful for measurements. For this, the Wigner--Weyl--Moyal formalism and the de Broglie--Madelung--Bohm theory will be predominant in our construction. 

Ultimately, we want to show that once the statistical arguments are adopted properly within a hydrodynamic interpretation, a kinetic theory and certain thermodynamic variables can be defined exactly for a quadratic system in question. What is more profound is that it is the underlying symplectic, classical dynamics that guides the quantum evolution of the squeezed coherent states and the corresponding thermodynamic variables.

In order to achieve this, we start Section~(\ref{sec:Preliminaries}) by providing some preliminaries. We give a brief summary of the de Broglie--Madelung--Bohm formalism. We then introduce a symplectic phase space for convenience as it is the starting point for the investigation of generic squeezed coherent states. Next, we briefly summarize Littlejohn's construction \cite{Littlejohn:1985} on the squeezed coherent states. The preliminaries section ends with the introduction of the Wigner function and its associated covariance matrix. Those will be important for identifying statistical and thermodynamical concepts.

In Section~(\ref{sec:A hydrodynamic interpretation for generalized squeezed coherent states}), we provide the main body of our own contribution on hydrodynamic interpretation of the generic squeezed coherent states. For this, in Section~(\ref{sec:Foundations of the construction}), we provide the foundations of our construction. We start with the polar decomposition of the Gaussian squeezed coherent state wave function that is exact. This allows us to decompose the Schr{\"{o}}dinger equation into pure real and imaginary parts which is the starting point of the de Broglie--Madelung--Bohm approach. Then, we present the associated Wigner function and identify its  covariance matrix in Section~(\ref{sec:Wigner function and the covariance matrix}). In the mean time, we keep track of the classical phase space shears, magnifications and rotations in order to identify which portion is responsible for which quantum statistical phenomenon in phase space. 

In Section~(\ref{sec:Probability distributions and entropies}), motivated by Moyal's statistical approach \cite{Moyal:1949}, we identify three types of probabilities: (i) a joint distribution, i.e., the Wigner function, which is defined on the entire phase space; (ii) a marginal distribution on position space; (iii) a conditional momentum distribution shaped by the post--selection of the positions. Note that all of the probability distributions listed above take Gaussian forms as the wave function in question is also Gaussian. We then make use of the Wigner distribution in order to define a phase space Shannon entropy. As the coherent states are known to be minimum uncertainty states, our phase space entropy takes an extremum value throughout the evolution of the system. This allows us to study equilibrium thermodynamics in a dynamical sense within a hydrodynamic interpretation.

Before introducing the full thermodynamic analysis, we derive the Fokker--Planck equation for the probability densities in Section~(\ref{sec:Fokker-Planck equation, probability fluxes and the continuity equation}). We show that it is not only the marginal position distribution that satisfies a continuity equation but all probability distributions. We discuss the probability flux related to the rotational degrees of freedom in addition to the one of the linear flow. Note that those Fokker--Planck equations are applicable for a dynamical situation unlike the standard case which was originally derived for stationary scenarios. 

We return back to the thermodynamic analysis in Section~(\ref{sec:Thermodynamic variables}), in which we start by presenting the analogy between the quantum phase space entropy and the Sackur--Tetrode entropy that was originally derived for the classical kinetic theory. Then, we follow Sonego's definitions in \cite{Sonego:1991} in order to obtain a quantum pressure and a quantum temperature for the squeezed coherent states. Next, we discuss the internal energy and its kinetic part that takes a similar form as in the classical kinetic theory. We show its relation to the quantum potential that is the key element of the de Broglie--Madelung--Bohm approach. Namely, we demonstrate that the quantum potential represents the kinetic internal energy of the system and the fluctuations around it at equilibrium. We also suggest a quantum virial relation which associates the conditional kinetic energy to a quantum potential energy term sourced solely by the quantum potential. In the end, we provide the link between the internal kinetic energy and the Maslov index defined for the symplectic paths. Essentially, we show that the quantum kinetic internal energy of a system is linked to the fractional Fourier transformations of the corresponding classical trajectories even if the system in question does not have periodic orbits. 

Finally, in Section~(\ref{sec:Summary and conclusion}), we provide a summary of our investigation, in addition to discussions regarding the extension of the domain of applicability of the current results.

\section{Preliminaries}\label{sec:Preliminaries}
\subsection{The de Broglie--Madelung--Bohm approach}\label{sec:de Broglie-Bohm approach}
There exists a correspondence between the paths taken by quantum particles and the paths taken by classical particles within the trajectory approach of the de Broglie--Madelung--Bohm \cite{DeBroglie:1925,DeBroglie:1927,Madelung:1927,Bohm:1952_I,Bohm:1952_II}. Depending on the interpretation, those trajectories either reflect a physical, tractable trajectory of a particle or an effective, mean stream--line trajectory of an ensemble of particles. The cost that has to be paid in return is the introduction of hidden variables to the theory whose existence has been debated in the literature many times.

Let us now introduce the summary of the causal theory in its original version.  For this, we will assume that there exists a particle with mass $m$ to which a complex wave function is assigned in its polar form
\begin{eqnarray}\label{eq:psi_polar}
\psi=R\exp{\left(\frac{i\mathcal{S}}{\hbar}\right)},
\end{eqnarray}
where $\hbar=h/(2\pi)$ with $h$ being the Planck's constant, $R=R\left(\mathbf{q},t\right)$ is a real amplitude and $\mathcal{S}=\mathcal{S}\left(\mathbf{q},t\right)$ is a real phase function. In the causal interpretation, the wave function satisfies the Schr{\"{o}}dinger equation in the following form
\begin{eqnarray}\label{eq:psi_schr_causal}
i\hbar \frac{\partial \psi}{\partial t}=\left(-\frac{\hbar ^2}{2m}\nabla ^2 +V\left(\mathbf{q},t\right)\right)\psi,
\end{eqnarray}
where $\boldsymbol{\nabla}={\partial}/{\partial \mathbf{q}}$ and $V\left(\mathbf{q},t\right)$ is the classical potential. In general, $V$ is a generic function of positions and it has no momentum dependence. Substitution of the wave function in its polar form, eq.~(\ref{eq:psi_polar}), into the Schr{\"{o}}dinger equation given in the form in eq.~(\ref{eq:psi_schr_causal}), results in a complex equation. Its pure imaginary and pure real parts are written respectively as
\begin{eqnarray}\label{eq:Sch_cpmlx_real}
\frac{\partial R}{\partial t}&=&-\frac{1}{2m}\left[R\,\nabla ^2 \mathcal{S}+2\boldsymbol{\nabla} R\cdot\boldsymbol{\nabla}\mathcal{S}\right]\label{eq:Sch_cmplx},\\
\frac{\partial \mathcal{S}}{\partial t}&=&-\left[\frac{\left(\boldsymbol{\nabla} \mathcal{S}\right)^2}{2m}+V\left(\mathbf{q},t\right)-\frac{\hbar ^2}{2m}\frac{\nabla ^2R}{R}\right].\label{eq:Sch_real}
\end{eqnarray}
The equations above have been interpreted in a hyrodynamic realm due to two main reasons. Firstly, one can define a flux--like term, $\mathbf{\tilde{j}}$, that is associated with a wave function,
\begin{eqnarray}\label{eq:flux_deBroglie-Madelung-Bohm}
\mathbf{\tilde{j}}= \frac{\hbar}{2mi}\left(\psi^*\boldsymbol{\nabla}\psi-\psi\boldsymbol{\nabla}\psi^*\right)=R^2\frac{\boldsymbol{\nabla}\mathcal{S}}{m},
\end{eqnarray}
where a probability density is defined through $\rho=\left[R\left(\mathbf{q},t\right)\right]^2$ and a velocity term is given by $\mathbf{\tilde{v}}={\boldsymbol{\nabla}\mathcal{S}}/{m}$. In that case $\mathbf{\tilde{j}}=\rho\mathbf{\tilde{v}}$ holds.
Then, the imaginary part of the Schr{\"{o}}dinger equation, eq.~(\ref{eq:Sch_cmplx}), can be viewed as a continuity equation,
\begin{eqnarray}\label{eq:cont_deBroglie-Madelung-Bohm}
\frac{\partial \rho}{\partial t}&+&\boldsymbol{\nabla}. \left(\rho \mathbf{\tilde{v}}\right)=0.
\end{eqnarray}

Secondly, when $\hbar \rightarrow 0$, the real part of the Schr{\"{o}}dinger equation,  (\ref{eq:Sch_real}), gives the Hamilton--Jacobi equation of the classical mechanics, i.e.,
\begin{eqnarray}
-\frac{\partial \mathcal{S}}{\partial t}=\frac{\left(\boldsymbol{\nabla} \mathcal{S}\right)^2}{2m}+V\left(\mathbf{q},t\right)=H,\label{eq:Ham-Jac}
\end{eqnarray}
with $\mathcal{S}$ playing the role of the action functional.

In the quantum case, the term
\begin{eqnarray}\label{eq:quantum_potential_original}
Q\left(\mathbf{q},t\right)=-\frac{\hbar ^2}{2m}\frac{\nabla ^2R}{R}
\end{eqnarray}
is non--zero and it appears in the quantum Hamilton--Jacobi equation in the same form as the classical potential, $V\left(\mathbf{q},t\right)$, does. That is why it is known as the \textit{quantum potential} which is responsible for the quantum behaviour of the given particles. The equation of motion now follows as
\begin{eqnarray}\label{eq:eom_deBroglie-Madelung-Bohm}
\frac{d\mathbf{\tilde{p}}}{dt}=-\bm{\nabla}\left(V+Q\right),
\end{eqnarray}
where the momentum $\mathbf{\tilde{p}}=m\mathbf{\tilde{v}}$ has different conceptualizations depending on the chosen interpretation. For instance, it might correspond to a single particle momentum within the de Broglie--Bohm theory. Whereas, in Madelung's hydrodynamic interpretation \cite{Madelung:1927} it corresponds to an effective momentum associated with an irrotational continuous ``fluid'' of particles.

Likewise, the quantum potential has different interpretations as well. According to Madelung, for example, the quantum potential is attributed to some quantum internal forces of a fluid. Alternatively, within a thermodynamic interpretation, the quantum potential can be related to the averaged kinetic energy of the quantum particles, the temperature of the corresponding system and the thermal vacuum energy \cite{Grossing:2009}. Similarly, it can be interpreted as the internal energy of a system \cite{Dennis:2015}. In certain investigations, quantum potential acts as an agent  that allows the interchange of information between systems \cite{Bohm:1984,Bohm:1995}. In the quantum cosmological realm, Bohmian interpretation and the quantum potential can even be related it to the dark energy problem \cite{Gonzalez:2012}. 

In summary, quantum potential allows one to identify the physical phenomena behind the quantum behaviour of a system. In Section~(\ref{sec:Internal energy, quantum potential and a conditional virial relation}), we will derive it for squeezed coherent states of time dependent systems in $n$--dimensions and we will interpret it thermodynamically similar to the ones in \cite{Grossing:2009,Dennis:2015}. 

\subsection{Symplectic phase space of classical orbits and the quadratic Hamiltonians }\label{sec:Symplectic phase space of quadratic Hamiltonians}
Let us now set up the phase space of a classical system by defining positions $q^a \in \mathbb{R}^n$ and momenta $p_a \in \mathbb{R}^n$ with $\{a, b\}=\{1...n\}$ as independent variables. We will consider only those Hamiltonians that are homogeneous quadratic functions of $\mathbf{q}$'s and $\mathbf{p}$'s which will be interpreted as canonical phase space coordinates.

For a classical system, consider a $2n$--dimensional phase space $N(\mathbb{R}^{2n},\omega)$ that is endowed with a symplectic form $\omega$ and Darboux coordinates $z^i= {(q^a, \,p_b)^{\intercal}}$. Here, $\{i, j\}=\{1...2n\}$ and $^{\intercal}$ refers to the transpose operator.

We write the Poisson bracket of two functions $f$ and $g$ as
\begin{eqnarray}\label{eq:Poisson_funcs}
\{f,g\}=\frac{\partial f}{\partial z^i}\omega ^{ij}\frac{\partial g}{\partial z^j},
\end{eqnarray}
where $\bm{\omega}$ is called the \textit{fundamental symplectic matrix}. 
It is defined through
\begin{eqnarray}\label{eq:Poisson_z}
\{z^i,z^j\}=\omega ^{ij}, \qquad
\omega ^{ij}=
\left[
\begin{array}{c|c}
\mathbf{0_n} & \, \, \mathbf{I_n} \\
\hline
\mathbf{-I_n} & \, \, \mathbf{0_n}
\end{array}
\right],
\end{eqnarray}
where $\mathbf{I_n}$ and $\mathbf{0_n}$ are $n$--dimensional identity and zero matrices, respectively. 

The matrix $\bm{\omega}$ satisfies
\begin{eqnarray}\label{eq:Omega_prop}
\bm{\omega}^{\intercal}=\bm{\omega}^{-1}=-\bm{\omega},\qquad \bm{\omega}^2=-\mathbf{I_{2n}},\qquad \rm{det}\,\bm{\omega}=1.\nonumber \\
\end{eqnarray}
Here, the inverse operator is denoted by $^{-1}$  and for the determinant of a matrix, we use ``${\rm{det}}$''. 
Given this, the symplectic two form acting on two arbitrary phase space vectors $\mathbf{z}$ and $\mathbf{z'}$ can be written as
\begin{eqnarray}\label{eq:symp_form}
\omega\left(\mathbf{z},\mathbf{z'}\right)=\mathbf{z}^\intercal\bm{\omega}^{-1}\mathbf{z'}=\mathbf{p}^\intercal\,\mathbf{q'}-\mathbf{q}^\intercal\mathbf{p'}.
\end{eqnarray}

We will now choose the Hamiltonian function, $H$, to be time dependent, i.e., $H=H\left(\mathbf{z},t\right)$ with $t\in \mathbb{R}$ and quadratic in $\mathbf{z}$. Then, $H$ is closed under the Poisson bracket (\ref{eq:Poisson_funcs}) and thus form a Lie algebra. Let us denote the Lie operator corresponding to $H$ as $\hat{\mathcal{L}}_{H}\left[\centerdot \right]=- \{ H, \centerdot \big\}$,
 which has a $2n\times 2n$ Hamiltonian matrix representation that we will denote by
\begin{eqnarray}\label{eq:Lie_matrix_Newtonian}
\mathbf{L}_{\mathbf{H}}=
\left[
\begin{array}{c|c}
\mathbf{b}^{\intercal}(t) & \, \, \mathbf{c}(t) \\
\hline
 -\mathbf{a}(t) & \, \,  -\mathbf{b}(t)
\end{array}
\right],
\end{eqnarray}
where $\mathbf{a}=\mathbf{a}^\intercal,\,\mathbf{c}=\mathbf{c}^\intercal$ and $\mathbf{b}$ are all $n\times n$ dimensional, time dependent, arbitrary matrices with $\mathbf{a},\,\mathbf{c}> \mathbf{0_n}$ and $\mathbf{a}\mathbf{c}-\mathbf{b}^2 > \mathbf{0_n}$. We will denote the set of all $2n\times 2n$ Hamiltonian matrices by ${h}:=\{\mathbf{L}_{\mathbf{H}}\in \mathbb{R}^{2n\times 2n}|\left(\bm{\omega}\mathbf{L}_{\mathbf{H}}\right)^\intercal=\left(\bm{\omega}\mathbf{L}_{\mathbf{H}}\right)\}.$

Now we write the Hamiltonian function as
\begin{eqnarray}\label{eq:H_cl}
H\left(\mathbf{z},t\right)= \frac{1}{2}\mathbf{z}^{\intercal}\bm{\omega}^\intercal\mathbf{L}_{\mathbf{H}}\mathbf{z},
\end{eqnarray}
such that the Hamiltonian equations take the form
\begin{eqnarray}\label{eq:Hamilton_eqs_z}
\hat{\mathcal{L}}_{H}\left[{z^i}\right]=-\{H,z^i\}=\omega ^{ij}\frac{\partial H}{\partial z^j},
\end{eqnarray}
or they are simply written as,
\begin{eqnarray}\label{eq:z_dot}
\frac{d\mathbf{z}}{ {dt}}= \mathbf{L}_{\mathbf{H}}\mathbf{z}.
\end{eqnarray}
Given the initial conditions $\mathbf{z_0}$ and $t_0=0$, the solution of eq.~(\ref{eq:z_dot}) is given by
\begin{eqnarray}\label{eq:S}
\mathbf{z}=\mathbf{S} {\left(t\right)}\mathbf{z_0},
\end{eqnarray}
where $\mathbf{S}$ is obtained by taking the exponential map of the Hamiltonian matrix $\mathbf{L}_{\mathbf{H}}$. Therefore, $\mathbf{S}$ is a $2n\times 2n$ symplectic matrix satisfying
\begin{eqnarray}\label{eq:Symplectic_S}
\mathbf{S}^{\intercal}\,\bm{\omega}\,\mathbf{S}=\bm{\omega}, \qquad \rm{det}\,\mathbf{S}=1.
\end{eqnarray}
Note that due to eqs.~(\ref{eq:z_dot}) and (\ref{eq:S}), $\mathbf{S}$ also follows the Hamiltonian flow, such that, 
\begin{eqnarray}\label{eq:S_dot}
\frac{d\mathbf{S}}{ {dt}}= \mathbf{L}_{\mathbf{H}}\mathbf{S},
\end{eqnarray}
holds for the initial conditions $\mathbf{S_0}=\mathbf{I_{2n}}$.
Let us write this matrix in the block form
\begin{eqnarray}\label{eq:symp_matrix}
\mathbf{S}(t)=
\left[
\begin{array}{c|c}
\mathbf{A}(t) & \, \, \mathbf{B}(t) \\
\hline
\mathbf{C}(t) & \, \, \mathbf{D}(t)
\end{array}
\right],
\end{eqnarray}
where $\mathbf{A},\,\mathbf{B},\,\mathbf{C}$ and $\mathbf{D}$ are all $n\times n$ matrices satisfying
\begin{eqnarray}
&&\mathbf{A}^\intercal\mathbf{C},\, \mathbf{B}^\intercal\mathbf{D}, \, \mathbf{A}\mathbf{B}^\intercal, \,
\mathbf{C}\mathbf{D}^\intercal \,\Rightarrow \rm{symmetric},\label{eq:Symp_cond_symm}\\
&&\mathbf{A}^\intercal\mathbf{D}-\mathbf{C}^\intercal\mathbf{B}=\mathbf{I_n},
\qquad {\rm{and}} \qquad
\mathbf{A}\mathbf{D}^\intercal-\mathbf{B}\mathbf{C}^\intercal=\mathbf{I_n},\label{eq:Symp_cond_Id}
\end{eqnarray}
due to the symplecticity conditions (\ref{eq:Symplectic_S}). Those matrices form the symplectic group $Sp\left(2n,\mathbb{R}\right)$ which has crucial importance for classical quadratic systems and their quantization. We will denote the set of all real $2n\times 2n$ symplectic matrices as
${s}:=\{\mathbf{S}\in \mathbb{R}^{2n\times 2n}|\mathbf{S}^{\intercal}\,\bm{\omega}\,\mathbf{S}=\bm{\omega}\}.$

In order to understand and identify the effect of a linear symplectic transformation on the evolution of phase space variables, one can use certain techniques to decompose symplectic matrices into its submatrices. However, not all of those decompositions are unique. On the other hand, Iwasawa showed that any symplectic matrix belonging to $Sp(2, \mathbb{R})$ can be decomposed uniquely into its nilpotent subgroup,  an abelian subgroup and a maximally compact subgroup \cite{Iwasawa:1949}. Those subgroups correspond to shearing, magnification and rotation effects on the phase space coordinates respectively. In the optics community, for example, those matrices represent lenses, magnifiers and fractional Fourier transformers.

Later, the Iwasawa decomposition was generalized to higher order symplectic matrices in which case the matrices responsible for the magnification effect do not form a group. Therefore, one refers to it as a factorization of the symplectic matrix or a modified--Iwasawa decomposition. It is given as \cite{Arvind:1995,Wolf:2004}
\begin{eqnarray}\label{eq:Iwasawa}
\mathbf{S}=
\left[
\begin{array}{c|c}
\mathbf{A} & \mathbf{B} \\
\hline
\mathbf{C} & \mathbf{D}
\end{array}
\right]
&=&\left[
\begin{array}{c|c}
\mathbf{I_2} & \mathbf{0_2} \\
\hline
\mathbf{-g} & \mathbf{I_2}
\end{array}
\right]
\left[
\begin{array}{c|c}
\mathbf{s} & \mathbf{0_2} \\
\hline
\mathbf{0_2} & \mathbf{s^{-1}}
\end{array}
\right]
\left[
\begin{array}{c|c}
\rm{Re}\mathbf{u} & \rm{Im}\mathbf{u} \\
\hline
-\rm{Im}\mathbf{u} & \rm{Re}\mathbf{u}
\end{array}
\right]\nonumber\\
\nonumber\\
&=&\qquad\mathbf{l(g)}\qquad \qquad \mathbf{m(s)}\qquad \qquad \,\,\,\, \mathbf{f(u)} 
\end{eqnarray}
where $\mathbf{l(g)}$ represents the shearing or lensing in phase space, $\mathbf{m(s)}$ represents magnifications and $\mathbf{f(u)}$, being a fractional Fourier transformer, represents rotation--like effects. Here, the $n\times n$ matrices that appear in (\ref{eq:Iwasawa}) are given in terms of the sub--blocks of the symplectic matrix, $\mathbf{S}$, as
\begin{eqnarray}\label{eq:Iwasawa_submatrices}
\mathbf{g}&=&-\left(\mathbf{C}\mathbf{A}^{\intercal}+\mathbf{D}\mathbf{B}^{\intercal}\right)\left(\mathbf{A}\mathbf{A}^{\intercal}+\mathbf{B}\mathbf{B}^{\intercal}\right)^{-1}=\mathbf{g}^{\intercal},\nonumber\\
\mathbf{s}&=&\left(\mathbf{A}\mathbf{A}^{\intercal}+\mathbf{B}\mathbf{B}^{\intercal}\right)^{1/2}=\mathbf{s}^{\intercal},\nonumber\\
\mathbf{u}&=&\left(\mathbf{A}\mathbf{A}^{\intercal}+\mathbf{B}\mathbf{B}^{\intercal}\right)^{-1/2}\left(\mathbf{A}+i\mathbf{B}\right)\in U(n).
\end{eqnarray}
In the following sections we will see that the modified Iwasawa factorization is useful in identifying how the different factors of the classical phase space evolution find their correspondences in the quantum mechanical evolution. 

In the next section, we re--present the squeezed coherent state wave function which is based on Littlejohn's construction \cite{Littlejohn:1985} and that is the central object of the current work.

\subsection{Generic squeezed coherent states}\label{sec:Squeezed coherent states}
Let us consider the quantum counterparts of the classical positions and momenta, represented in the position space as,  $\hat{\mathbf{q}}=\left(\hat{q}_1, \hat{q}_2,...,\hat{q}_n \right)^\intercal$ and $\hat{\mathbf{p}}=\left(\hat{p}_1, \hat{p}_2,...,\hat{p}_n \right)^\intercal$, respectively. 
 
They operate on a function ${f}$ as
\begin{eqnarray}
\hat{\mathbf{q}}\left[{f}\right]= {f} \,\cdot \, \mathbf{q},\qquad {\rm{and}} \qquad
\hat{\mathbf{p}}\left[{f}\right]=-i\hbar\frac{\partial\, {f}}{\partial \mathbf{q}},
\end{eqnarray}
where the $``\,\cdot\,"$ denotes the standard multiplication. They satisfy the Heisenberg commutation rule $\left[\hat{{p}}_k,\, \hat{{q}}_j\right]=-i\hbar\delta _{kj},$
 where $\delta _{kj}$ is the Kronecker delta.
Likewise, the quantum Hamiltonian operator which is the counterpart of the quadratic classical Hamiltonian given in eq.~(\ref{eq:H_cl}) is written as
\begin{eqnarray}\label{eq:Hamil_Op}
\hat{H}(t)= \frac{1}{2}\hat{\mathbf{z}}^\dag\bm{\omega}^\intercal \mathbf{L}_{\mathbf{H}}(t)\hat{\mathbf{z}}\qquad \rm{with}\qquad \hat{\mathbf{z}}= 
\left[
\begin{array}{c}
\hat{\mathbf{q}} \\
\hat{\mathbf{p}}
\end{array}
\right],
\end{eqnarray}
where ``$\dag$'' denotes the conjugate transpose and $\hat{\mathbf{z}}$ is the quantum counterpart of the phase space vector ${\mathbf{z}}$. For the quadratic system given in eq.~(\ref{eq:Hamil_Op}), the expectation value of $\hat{\mathbf{z}}$ follows the underlying classical trajectory due to the Ehrenfest Theorem. This means $\left\langle \hat{\mathbf{z}}\right\rangle = \left\langle \Psi \right| \hat{\mathbf{z}}\left| \Psi \right\rangle = {\mathbf{z}} $, where $\left| \Psi \right\rangle$ is the state which has a wave function $\Psi$. 

In order to define and study the phase space evolution of the wave functions of the squeezed coherent states, one considers two sets of operations \cite{Littlejohn:1985}: (i) translations given by the Weyl--Heisenberg operators, $\hat{T}$, and (ii) the squeezings generated by the metaplectic operators, $\hat{M}(\mathbf{S})$. The latter are associated with the symplectic matrices $\mathbf{S}$ that guide the underlying classical evolution.

It is known that there exists a unitary operator  $ \hat{U}\left(t,\mathbf{z_0}\right)$, which incorporates both the action of translations and the squeezings in the phase space. This propagator satisfies the Schr{\"{o}}dinger equation just like the wave function itself,
\begin{eqnarray}\label{eq:Sch_U_prop}
i\hbar \frac{d\hat{U}}{dt}=\hat{H}\hat{U}\qquad \rm{with} \qquad \hat{U}\left(0,\mathbf{z_0}\right)=\rm{Identity}.
\end{eqnarray}

In order to find out how this propagator acts on an initial wave function which is centered at $\mathbf{z_0}$, one can indeed make use of the ground state, $\left| \mathbf{0}\right\rangle$, that is centered at $\mathbf{0}$. 

At this point, we refer to \ref{appendix:squeezed coherent states}, in which we give a brief summary of  the derivation of the squeezed coherent state wave function. Note that this derivation is mostly based on Littlejohn's construction \footnote{\ref{appendix:squeezed coherent states} involves only those points that are immediately relevant for us which does not give the full credit to the original paper.} \cite{Littlejohn:1985}. 
For example, substituting eq.~(\ref{eq:U_prop}) into eq.~(\ref{eq:psi_U_psi_0}) with $\left| \psi_0\right\rangle=\hat{T} \left(\mathbf{z_0}\right)\left| \mathbf{0}\right\rangle$ gives
\begin{eqnarray}
\left| \psi\right\rangle=
\hat{U}\left(t,\mathbf{z_0}\right)\left| \psi_0\right\rangle
=\exp{\left(\frac{i\gamma (t)}{\hbar}\right)}\hat{T}\left( \mathbf{z} (t)\right)\hat{M}\left(\mathbf{S}(t)\right)\left| \mathbf{0}\right\rangle,
\end{eqnarray}
where the phase function $\gamma (t)$ is given by eq.~(\ref{eq:Phase_Littlejohn}).
This means that regardless of what the initial state is, one can make use of a fiducial, ground state in order to obtain the final state. This might seem counter intuitive at a first glance. However, note that the information about the initial phase space vector is already included in the symplectic matrix $\mathbf{S}$ and its quantum counterpart $\hat{M}(\mathbf{S})$.

Finally, since the ground state wave function is represented in position space by
\begin{eqnarray}\label{eq:psi_ground}
\psi_{\left| \mathbf{0}\right\rangle}=\frac{1}{\left(\pi \hbar\right)^{n/4}}\exp{\left(-\frac{\mathbf{q}^\intercal\mathbf{q}}{2\hbar}\right)},
\end{eqnarray}
one can obtain the matrix representation of the squeezed coherent state wave function defined at time $t$ and centered at the phase space point $\langle \hat{\mathbf{z}} \rangle=\left(\langle \hat{\mathbf{q}}\rangle, \langle \hat{\mathbf{p}}\rangle \right)^\intercal$ in its exact form as \cite{Littlejohn:1985}

\begin{eqnarray}\label{eq:psi_seq_coh}
\psi &=&\frac{1}{\left(\pi \hbar\right)^{n/4}}\exp{\left(\frac{i\gamma (t)}{\hbar}\right)}\hat{T} \left( \mathbf{z} (t)\right)\frac{1}{\sqrt{\rm{det}\left(\mathbf{A}+i\mathbf{B}\right)}}\exp{\left[\frac{i}{2\hbar}\left(\mathbf{q}^\intercal \mathbf{\Gamma}\mathbf{q}\right)\right]}\nonumber \\
&=&\frac{1}{\left(\pi \hbar\right)^{n/4}}\frac{1}{\sqrt{\rm{det}\left(\mathbf{A}+i\mathbf{B}\right)}}\exp{\left[\frac{i} {\hbar}\left(\gamma(t)+\langle \hat{\mathbf{p}}\rangle^\intercal\mathbf{q}-\frac{\langle \hat{\mathbf{p}}\rangle^\intercal\langle \hat{\mathbf{q}}\rangle}{2}+\frac{1}{2}\left(\mathbf{q}-\langle\hat{\mathbf{q}}\rangle\right)^\intercal \mathbf{\Gamma}\left(\mathbf{q}-\langle \hat{\mathbf{q}}\rangle\right)\right)\right]},\nonumber\\
\end{eqnarray}

where $\mathbf{\Gamma}=\left(\mathbf{C}+i\mathbf{D}\right)\left(\mathbf{A}+i\mathbf{B}\right)^{-1}$. 

Our aim in the current work is to find the hydrodynamic interpretation of a system represented by the wave function $\psi$ given in eq.~(\ref{eq:psi_seq_coh}). We will essentially study this system within the realm of statistical mechanics. Therefore, quantum mechanical distribution functions are essential for our investigation. Thus, we will now introduce certain concepts that were previously introduced into the literature and which will be useful in our statistical construction.

\subsection{Wigner function, Wigner ellipsoid and the covariance matrix}\label{sec:Wigner ellipse and the Lewis-Riesenfeld invariant}
For a given state $\left| \psi\right\rangle$, one can associate a function, known as the Wigner function  $W=W(\mathbf{q},\mathbf{p})$, which is the Weyl symbol of the projection operator $\left| \psi\right\rangle \left\langle \psi \right|$. In the context of quantum mechanics, it was introduced into the literature by Wigner \cite{Wigner:1932} as a quasi--probability distribution. Wigner function has similar properties to the phase space distribution function of classical mechanics which is preserved via the Liouville equation. For a state, $\left| \psi\right\rangle$, in $\mathbf{q}$--representation it is given by \cite{Wigner:1932}
\begin{eqnarray}\label{eq:Wigner}
W=\frac{1}{\left(2\pi \hbar\right)^{n}}\int \psi\left(\mathbf{q}+\frac{\mathbf{x}}{2}\right)\psi ^{*}\left(\mathbf{q}-\frac{\mathbf{x}}{2}\right) \exp{\left(-\frac{i}{\hbar}\mathbf{p}\mathbf{x}\right)}d\mathbf{x}.
\end{eqnarray}

Wigner function has very nice transformation properties. Let us consider the Weyl--Heisenberg operator, $\hat{T}( \mathbf{z'})$, and the metaplectic operator, $\hat{M}\left(\mathbf{S}\right)$, introduced in the previous section. It is known that if  $W(\mathbf{z})$ is the Wigner function of a state $\left| \psi\right\rangle$, then, $W(\mathbf{z}- \mathbf{z'})$ is the one of the translated state $\hat{T}(\mathbf{z'})\left| \psi\right\rangle$ \cite{Littlejohn:1985}. In addition, the transformation of the Wigner function under a symplectic transformation gives $W(\mathbf{S^{-1}z})$. Namely, if  $W(\mathbf{z})$ is the Wigner function of a state $\left| \psi\right\rangle$, then, $W(\mathbf{S^{-1}z})$ is the Wigner function of the state $\hat{M}\left(\mathbf{S}\right)\left| \psi\right\rangle$.  
This results in 
\begin{eqnarray}
W(\mathbf{z},t)=W(\mathbf{S^{-1}z},0),
\end{eqnarray}
such that the Wigner function is invariant throughout the evolution of the system.

The expectation values of quantum operators can be obtained via the Wigner function with an integral transform similar to the classical phase space averaging.
For an operator, $\hat{F}$, for example, its expectation value is obtained through
\begin{equation}\label{eq:Wigner_ave}
\left\langle \psi \right|\hat{F}\left| \psi\right\rangle= \int {d\mathbf{z}}\,W(\mathbf{z})f(\mathbf{z}),
\end{equation}
where the phase space function $f(\mathbf{z})$ is the Weyl symbol corresponding to the operator $\hat{F}$. Also note that the distribution $W(\mathbf{z})$ is normalized, i.e., 
\begin{eqnarray}
\int {d\mathbf{z}}\,W(\mathbf{z})=1.
\end{eqnarray}

For a Wigner function, which is centered at $\left\langle \hat{\mathbf{z}}\right \rangle$, the first order moments can be calculated via
\begin{equation}\label{eq:ave_z}
\left\langle \hat{\mathbf{z}} \right \rangle= \int {d\mathbf{z}}\,W(\mathbf{z})\mathbf{z}.
\end{equation}
Moreover, a covariance matrix, $\mathbf{\Sigma}$, can be calculated in a similar manner via the second moments of the Wigner function via
\begin{equation}\label{eq:Cov_Wigner}
\mathbf{\Sigma}_{\alpha \beta} = \int {d\mathbf{z}}\,W(\mathbf{z})\mathbf{z}_\alpha \mathbf{z}_\beta.
\end{equation}
Note that the matrix $\mathbf{\Sigma}$ is symmetric and non-negative.
By using eq.~(\ref{eq:Cov_Wigner}), one can show for Gaussian states that Wigner function in eq.~(\ref{eq:Wigner}) also takes a Gaussian form \cite{Bastiaans:1979, Littlejohn:1985, Yeh:1993}
\begin{eqnarray}\label{eq:Wigner_Gaussian}
W=W(\mathbf{z},t)
=\frac{1}{\left(\pi \hbar\right)^n}\exp{\left\{-\frac{1}{\hbar}\left(\mathbf{z}-\left\langle \hat{\mathbf{z}} \right \rangle\right)^\intercal \mathbf{W}\left(\mathbf{z}-\left\langle \hat{\mathbf{z}} \right \rangle\right)\right\}},\qquad {\rm{where}} \qquad \mathbf{W}=\frac{\hbar}{2} \mathbf{\Sigma}^{-1}.
\end{eqnarray}

We know that for the squeezed coherent states, the first moments follow classical trajectories, i.e., $\langle \hat{\mathbf{z}} \rangle (t)=\mathbf{S}\,\langle \hat{\mathbf{z}}\rangle (0)$. Also the Wigner function being preserved in the phase space gives
\begin{eqnarray}\label{eq:Wigner_mat_evol}
\mathbf{W}(t)=\mathbf{S}^{-\intercal}\mathbf{W}(0)\mathbf{S}^{-1}, \,\, \rm{with} \,\, \rm{det}\mathbf{W}=1.
\end{eqnarray}
Therefore, for the squeezed coherent states, $\mathbf{W}$ is symplectic, symmetric and positive definite \cite{Bastiaans:1979}.
One can then choose, for example, \cite{Yeh:1993}
\begin{eqnarray}\label{eq:Wigner_ellipse}
\left(\mathbf{z}-\left\langle \hat{\mathbf{z}} \right \rangle\right)^\intercal. \mathbf{W}.\left(\mathbf{z}-\left\langle \hat{\mathbf{z}} \right \rangle\right)=1,
\end{eqnarray}
which defines the surface of an ellipsoid centered at $\left\langle \hat{\mathbf{z}} \right \rangle$. This means that both the surface and the center of the Wigner ellipsoid transforms rigidly throughout the evolution of the system \cite{Littlejohn:1985,Yeh:1993}. 

Moreover, as the invariants of a system are associated with its symmetries, one might wonder which quantum invariants are preserved under the symplectic symmetries of the classical system. Therefore, we include a discussion in \ref{appendix:Invariants}, for a curious reader, that summarizes the relationship of the Wigner ellipsoid with some classical and quantum mechanical invariants of linear systems.

In this section, we presented the preliminaries which are essential for our investigation. Those seemingly unrelated ingredients come together in the main body of our work in Section~(\ref{sec:A hydrodynamic interpretation for generalized squeezed coherent states}) in the following manner. We consider a quadratic system with a classical symplectic phase space dynamics as in Section~(\ref{sec:Symplectic phase space of quadratic Hamiltonians}). A quantum correspondence of this system is considered via Littlejohn's squeezed coherent state wave function that was shortly presented in Section~(\ref{sec:Squeezed coherent states}). Note that we would like to interpret this system as in the de Broglie--Madelung--Bohm approach which was summarized in Section~(\ref{sec:de Broglie-Bohm approach}). For our investigation, this is a hydrodynamic interpretation intertwined with certain statistical and thermodynamic concepts. Therefore, we will also make use of the definition of the Wigner function and the Wigner--Weyl--Moyal correspondence as summarized in Section~(\ref{sec:Wigner ellipse and the Lewis-Riesenfeld invariant}). 

\section{A hydrodynamic interpretation for generic squeezed coherent states}\label{sec:A hydrodynamic interpretation for generalized squeezed coherent states}
Previously, in Section (\ref{sec:de Broglie-Bohm approach}), we summarized the causal approach of the de Broglie--Madelung--Bohm theory. Let us recall that the entire formalism depends on a wave function being written on its polar form, i.e., $\psi=R\exp{\left(\frac{i\mathcal{S}}{\hbar}\right)}$ where $R$ and $\mathcal{S}$ are real functions. Therefore, in order to start our investigation, we need to transform the wave equation of the squeezed coherent state given in eq.~(\ref{eq:psi_seq_coh}) into its polar form first. This is what we present in the next section.
\subsection{Foundations of the construction}\label{sec:Foundations of the construction}
\subsubsection{Polar decomposition}\label{sec:Polar decomposition}
The Hamiltonians we consider here are in quadratic order, thus any choice of operator ordering will result in the same outcome. Let us choose the symmetric ordering and consider the following Hamiltonian operator 
\begin{eqnarray}\label{eq:Ham_gen}
\hat{H}=\frac{1}{2}\hat{\mathbf{q}}^\dag\mathbf{a}\hat{\mathbf{q}}+\frac{1}{2}\left(\hat{\mathbf{q}}^\dag\mathbf{b}\hat{\mathbf{p}}+\hat{\mathbf{p}}^\dag\mathbf{b}^\intercal\hat{\mathbf{q}}\right)+\frac{1}{2}\hat{\mathbf{p}}^\dag\mathbf{c}\hat{\mathbf{p}}.\nonumber \\
\end{eqnarray}
Now we write the Schr{\"{o}}dinger equation by using the generic Hamiltonian operator in eq.~(\ref{eq:Ham_gen}), so that,
\begin{align}\label{eq:psi_schr_gen}
i\hbar \frac{\partial \psi}{\partial t}=\left(\frac{1}{2}{\mathbf{q}}^\intercal\mathbf{a}{\mathbf{q}}-\frac{i\hbar}{2}\left[{\rm{Tr}}(\mathbf{b})+2\mathbf{q}^\intercal\mathbf{b}\boldsymbol{\nabla _{\mathbf{q}}}\right]-\frac{\hbar ^2}{2}\boldsymbol{\nabla_{\mathbf{q}}}^\intercal\,\mathbf{c}\boldsymbol{\nabla_{\mathbf{q}}}\right)\psi,
\end{align}
where
\begin{eqnarray}
\boldsymbol{\nabla _{\mathbf{q}}}=\frac{\partial}{\partial \mathbf{q}}=
\left[
\frac{\partial}{\partial q_1} \,
\frac{\partial}{\partial q_2} \,
...
\frac{\partial}{\partial q_n}
\right]^\intercal.
\end{eqnarray}
Then, for a wave function which is written in its polar form  (\ref{eq:psi_polar}), we obtain the pure imaginary and the pure real parts of the Schr{\"{o}}dinger eq.~(\ref{eq:psi_schr_gen}) respectively as
\begin{align}
\frac{\partial R}{\partial t}=-\frac{1}{2}\left[R\boldsymbol{\nabla_{\mathbf{q}}}^\intercal\mathbf{c} \boldsymbol{\nabla_{\mathbf{q}}} \mathcal{S} +\left(\boldsymbol{\nabla_{\mathbf{q}}} \mathcal{S}\right)^\intercal\mathbf{c}\boldsymbol{\nabla_{\mathbf{q}}} R +\left(\boldsymbol{\nabla_{\mathbf{q}}} R\right)^\intercal\mathbf{c}\boldsymbol{\nabla_{\mathbf{q}}} \mathcal{S}+2\mathbf{q}^\intercal\mathbf{b}\boldsymbol{\nabla_{\mathbf{q}}} R +{\rm{Tr}}(\mathbf{b})R\right], \label{eq:Sch_imaginary}
\end{align}
and,
\begin{align}
\frac{\partial \mathcal{S}}{\partial t}=-\left[\frac{1 }{2}\left(\boldsymbol{\nabla_{\mathbf{q}}} \mathcal{S}\right)^\intercal\mathbf{c}\boldsymbol{\nabla_{\mathbf{q}}} \mathcal{S}+\frac{1}{2}\mathbf{q}^\intercal\mathbf{a}\mathbf{q} -\frac{\hbar ^2}{2}\frac{1}{R}\boldsymbol{\nabla_{\mathbf{q}}}^\intercal\mathbf{c}\boldsymbol{\nabla_{\mathbf{q}}} R+\mathbf{q}^\intercal\mathbf{b}\boldsymbol{\nabla_{\mathbf{q}}} \mathcal{S}\right].
\label{eq:Sch_real_generic}
\end{align}
When we compare the real part of the Schr{\"{o}}dinger equation in the generic case, i.e., eq.~(\ref{eq:Sch_real_generic}), with the one of the original definition in eq.~(\ref{eq:Sch_real}), we realise that the general quantum potential for an $n$--dimensional system is
\begin{eqnarray}\label{eq:quant_pot_matrix}
Q=-\frac{\hbar ^2}{2}\frac{1}{R}\boldsymbol{\nabla_{\mathbf{q}}}^\intercal\mathbf{c}\boldsymbol{\nabla_{\mathbf{q}}} R.
\end{eqnarray}
This is the analogous expression for eq.~(\ref{eq:quantum_potential_original}).

In order to obtain the explicit form of the equation set~(\ref{eq:Sch_imaginary})-(\ref{eq:Sch_real_generic}) for generic squeezed coherent states, we need to write the wave function given in eq.~(\ref{eq:psi_seq_coh}), i.e.,

\begin{eqnarray}\label{eq:psi_seq_coh_2}
\psi=\frac{1}{\left(\pi \hbar\right)^{n/4}}\frac{1}{\sqrt{\rm{det}\left(\mathbf{A}+i\mathbf{B}\right)}}\exp{\left[\frac{i} {\hbar}\left(\gamma(t)+\langle \hat{\mathbf{p}}\rangle^\intercal\mathbf{q}-\frac{\langle \hat{\mathbf{p}}\rangle^\intercal\langle \hat{\mathbf{q}}\rangle}{2}+\frac{1}{2}\left(\mathbf{q}-\langle\hat{\mathbf{q}}\rangle\right)^\intercal \mathbf{\Gamma}\left(\mathbf{q}-\langle \hat{\mathbf{q}}\rangle\right)\right)\right]}
\end{eqnarray}

in its polar form. Note that it is not immediately obvious whether this is possible for a generic case due to the $\mathbf{\Gamma}=\left(\mathbf{C}+i\mathbf{D}\right)\left(\mathbf{A}+i\mathbf{B}\right)^{-1}$ term that appears in the exponent and the $\sqrt{{\rm{det}}\left(\mathbf{A}+i\mathbf{B}\right)}$ term that appears in the denominator. However, we will see that the modified Iwasawa factorization, eq.~(\ref{eq:Iwasawa}) of Section~(\ref{sec:Symplectic phase space of quadratic Hamiltonians}), will help us in identifying the pure real and the pure imaginary parts of the wave function of the squeezed coherent states.

Let us recall that the classical evolution of the system is governed by a symplectic matrix $\mathbf{S}$ as in eq.~(\ref{eq:symp_matrix}), whose action on the phase space can be factored into three effects: shearing, magnification and rotation. The latter, the fractional Fourier transformer part of the Iwasawa factorization, is represented by a matrix $\mathbf{f(u)}$ given by 
\begin{eqnarray}\label{eq:frac_Four_trans}
\mathbf{f(u)}=
\left[
\begin{array}{c|c}
\rm{Re}\mathbf{u} & \rm{Im}\mathbf{u} \\
\hline
-\rm{Im}\mathbf{u} & \rm{Re}\mathbf{u},
\end{array}
\right],
\end{eqnarray}
as we have already re--presented in Section~(\ref{sec:Symplectic phase space of quadratic Hamiltonians}).
Here, $\mathbf{u}$ is a unitary matrix given by
\begin{eqnarray}
\mathbf{u}=\left(\mathbf{A}\mathbf{A}^{\intercal}+\mathbf{B}\mathbf{B}^{\intercal}\right)^{-1/2}\left(\mathbf{A}+i\mathbf{B}\right).
\end{eqnarray}
Since unitary matrices can uniquely be written in the form $\mathbf{u}=\mathbf{\bar{u}}\mathbf{\tilde{u}}$ where $\mathbf{\bar{u}}\in$ SU(N) and $\mathbf{\tilde{u}}=\exp{\left(i\alpha/n\right)}\mathbf{I_n}$, we have $\rm{det}\mathbf{u}=\exp{\left(i\alpha\right)}$,
and
\begin{eqnarray}\label{eq:deter_AiB}
\sqrt{{\rm{det}}\left(\mathbf{A}+i\mathbf{B}\right)}=\exp{\left(i\alpha/2\right)}\sqrt{{\rm{det}}\,\mathbf{s}}.
\end{eqnarray}
Note that $\mathbf{s}=\left(\mathbf{A}\mathbf{A}^{\intercal}+\mathbf{B}\mathbf{B}^{\intercal}\right)^{1/2}=\mathbf{s}^\intercal$ is the matrix that is responsible for pure magnifications in phase space as denoted in eq.~(\ref{eq:Iwasawa_submatrices}). In that case, we have\footnote{A similar result can be found in \cite{deGosson:2019arxiv}.}
\begin{eqnarray}\label{eq:Gamma_lens_magn}
\mathbf{\Gamma}&=&\left(\mathbf{C}+i\mathbf{D}\right)\left(\mathbf{A}+i\mathbf{B}\right)^{-1}\nonumber\\
&=&\left(\mathbf{C}+i\mathbf{D}\right)\left(\mathbf{A}-i\mathbf{B}\right)^\intercal\left[\left(\mathbf{A}+i\mathbf{B}\right)\left(\mathbf{A}-i\mathbf{B}\right)^\intercal\right]^{-1}\nonumber\\
&=&\left(\mathbf{C}\mathbf{A}^\intercal+\mathbf{D}\mathbf{B}^\intercal+i\mathbf{I_n}\right)\mathbf{s}^{-2}\nonumber\\
&=&-\mathbf{g}+i\mathbf{s}^{-2},
\end{eqnarray}
where the third line in eq.~(\ref{eq:Gamma_lens_magn}) follows from the symplecticity conditions $\mathbf{A}\mathbf{B}^\intercal=\mathbf{B}\mathbf{A}^\intercal$ and $\mathbf{A}\mathbf{D}^\intercal-\mathbf{B}\mathbf{C}^\intercal=\mathbf{I_n}$ of $\mathbf{S}$ given in eqs.~(\ref{eq:Symp_cond_symm}) and (\ref{eq:Symp_cond_Id}). Note that $\mathbf{g}=-\left(\mathbf{C}\mathbf{A}^{\intercal}+\mathbf{D}\mathbf{B}^{\intercal}\right)\left(\mathbf{A}\mathbf{A}^{\intercal}+\mathbf{B}\mathbf{B}^{\intercal}\right)^{-1}=\mathbf{g}^{\intercal}$ appears in the Iwasawa factorization (\ref{eq:Iwasawa}) as part of the symplectic evolution that is responsible for the shearing effect in $\mathbf{p}$--space. In optics, for example, its associated symplectic matrix $\mathbf{l(g)}$ corresponds to the propagation through a thin lens.

Then, once we substitute the expressions (\ref{eq:deter_AiB}) and (\ref{eq:Gamma_lens_magn}) into eq.~(\ref{eq:psi_seq_coh_2}), we obtain the polar decomposition of the wave function of the squeezed coherent state. We find the real amplitude and the real phase function respectively as
\begin{align}
R=\frac{1}{\left(\pi \hbar\right)^{n/4}}\frac{1}{\sqrt{{\rm{det}}\mathbf{s}}}\exp{\left[-\frac{1}{2\hbar}\left(\mathbf{q}-\langle \hat{\mathbf{q}}\rangle\right)^\intercal \mathbf{s}^{-2} \left(\mathbf{q}-\langle \hat{\mathbf{q}}\rangle\right)\right]}\label{eq:ampl_sq_coh},
\end{align}
and,
\begin{align}
\mathcal{S}=\gamma(t)-\frac{\hbar\alpha}{2}+\langle\hat{\mathbf{p}}\rangle^\intercal\mathbf{q}-\frac{\langle\hat{\mathbf{p}}\rangle^\intercal\langle \hat{\mathbf{q}}\rangle}{2}-\frac{1}{2}\left(\mathbf{q}-\langle\hat{\mathbf{q}}\rangle\right)^\intercal \mathbf{g}\left(\mathbf{q}-\langle\hat{\mathbf{q}}\rangle\right)\label{eq:phase_sq_coh}.
\end{align}

It can clearly be seen that the real amplitude of the wave function is governed solely by the matrix $\mathbf{s}$ that appears in the symplectic magnification matrix $\mathbf{m(s)}$. On the other hand, the phase function is governed by the matrix $\mathbf{g}$ responsible for the shearing/lensing effect, $\mathbf{l(g)}$, adopted from the underlying classical phase space evolution.

In the next section, we will make use of the definitions reintroduced in Section~(\ref{sec:Wigner ellipse and the Lewis-Riesenfeld invariant}) in order to find a phase space distribution function and a covariance matrix associated with the squeezed coherent state wave function.

\subsubsection{Wigner function and the covariance matrix}\label{sec:Wigner function and the covariance matrix}
It is known that there exists no unique definition of a probability distribution in quantum mechanics.
The Wigner function, being the  Weyl symbol of the density operator, is one of the candidates to be chosen as the quasi--probability distribution of a phase space. The alternatives to the Wigner distribution are (i) the Husimi distribution \cite{Husimi:1940} which is sometimes referred to as the ``regularization'' of the Wigner distribution \cite{Combescure:2012}; (ii) the Glauber--Sudarshan distribution \cite{Glauber:1963,Sudarshan:1963} which was original derived for the coherent states.

Note that the Wigner function might take negative values within certain scenarios, whereas the Husimi distribution does not have this property. On the other hand, it is argued that \cite{Colomes:2015}, the Husimi distribution does not provide the correct charge and current densities for certain cases \footnote{One can refer to \cite{Kiesel:2013} for the comparison of different quasi--probability distributions in the literature. Also see \cite{Manko:2021} for the generic invertible maps of density operators onto probability distributions for a broader review.}. As the Wigner function is also non--negative for the Gaussian states, we choose the Wigner function to be used as a proper phase space distribution function in our current investigation.

In order to obtain the explicit form of the Wigner function as presented in Section~(\ref{sec:Wigner ellipse and the Lewis-Riesenfeld invariant}), we follow \cite{Bastiaans:1979} and \cite{Littlejohn:1985} by considering eq.~(\ref{eq:Gamma_lens_magn}). Then, for a Gaussian Wigner function,
\begin{eqnarray}\label{eq:Wignerfnc}
W=\frac{1}{\left(\pi \hbar \right)^n}\exp{\left[\frac{-1}{\hbar}\left({\mathbf{z}}-\left\langle \hat{\mathbf{z}} \right \rangle\right)^\intercal \mathbf{W}\left({\mathbf{z}}-\left\langle \hat{\mathbf{z}} \right \rangle\right)\right]},\nonumber \\
\end{eqnarray}
the Wigner matrix, $\mathbf{W}$, is a $2n\times 2n$ symplectic matrix that takes the form  
\begin{eqnarray}\label{eq:Wigner_matrix_coh}
\mathbf{W}=
\left[
\begin{array}{c|c}
\left(\mathbf{s}^{-2}+\mathbf{gs^2g}\right) & \, \, \mathbf{gs^2}) \\
\hline
\mathbf{s^2g} & \, \, \mathbf{s^2}
\end{array}
\right],
\end{eqnarray}
in our case.

As mentioned before, a covariance matrix can be obtained via the Wigner distribution by making use of the eq.~(\ref{eq:Cov_Wigner}) such that  $\boldsymbol{\Sigma}=\frac{\hbar}{2}\mathbf{W}^{-1}$. Then, we find the covariance matrix associated with the squeezed coherent state as
\begin{eqnarray}\label{eq:Covariance_matrix_coh}
\boldsymbol{\Sigma}&=&
\left[
\begin{array}{c|c}
\boldsymbol{\sigma_{qq}} & \, \, \boldsymbol{\sigma_{qp}} \\
\hline
\boldsymbol{\sigma_{pq}} & \, \, \boldsymbol{\sigma_{pp}}
\end{array}
\right]
=
\left[
\begin{array}{c|c}
\left\langle \mathbf{\hat{q}}^2 \right \rangle - \left\langle \mathbf{\hat{q}} \right \rangle^2& \, \, 
\left\langle \frac{\mathbf{\hat{q}\hat{p}}+\mathbf{\hat{p}\hat{q}}}{2}  \right \rangle - \left\langle \mathbf{\hat{q}} \right \rangle\left\langle \mathbf{\hat{p}} \right \rangle\\
\hline
\left\langle \frac{\mathbf{\hat{p}\hat{q}}+\mathbf{\hat{q}\hat{p}}}{2}  \right \rangle - \left\langle \mathbf{\hat{p}} \right \rangle\left\langle \mathbf{\hat{q}} \right \rangle & 
\, \, \left\langle \mathbf{\hat{p}}^2 \right \rangle - \left\langle \mathbf{\hat{p}} \right \rangle^2
\end{array}
\right]\nonumber \\
&=&
\frac{\hbar}{2}
\left[
\begin{array}{c|c}
\mathbf{s^2} & \, \, -\mathbf{s^2g} \\
\hline
-\mathbf{gs^2} & \, \, \left(\mathbf{s}^{-2}+\mathbf{gs^2g}\right)
\end{array}
\right].\nonumber
\\
\end{eqnarray}
Moreover, the invariance of the Wigner ellipsoid dictates that
\begin{eqnarray}
\mathbf{\Sigma}(t)=\mathbf{S}(t)\mathbf{\Sigma}(0)\mathbf{S}^{\intercal}(t).
\end{eqnarray}
Then, we get
\begin{eqnarray}\label{eq:Sigma_dot}
\frac{d\mathbf{\Sigma}}{dt}=\mathbf{L}_{\mathbf{H}}\mathbf{\Sigma}+\mathbf{\Sigma}\mathbf{L}_{\mathbf{H}}^{\intercal}, \qquad {\rm{with}} \qquad \mathbf{L}_{\mathbf{H}}=
\left[
\begin{array}{c|c}
\mathbf{b}^{\intercal}(t) & \, \, \mathbf{c}(t) \\
\hline
 -\mathbf{a}(t) & \, \,  -\mathbf{b}(t)
\end{array}
\right],
\end{eqnarray}
due to eq.~(\ref{eq:S_dot}). 

In order to find out how the sub--matrices evolve, we substitute the explicit form of $\mathbf{\Sigma}$ in eq.~(\ref{eq:Covariance_matrix_coh}), into its  time evolution above. Then we obtain,
\begin{eqnarray}
\frac{d\mathbf{\mathbf{s}^{-2}}(t)}{dt}&=&-\mathbf{s}^{-2}\mathbf{b}^\intercal-\mathbf{b}\mathbf{s}^{-2}+\mathbf{s}^{-2}\mathbf{c}\mathbf{g}+\mathbf{g}\mathbf{c}\mathbf{s}^{-2},\nonumber \\
\label{eq:s-2_dot}\\
\frac{d\mathbf{g}(t)}{dt}&=&\mathbf{a}-\mathbf{bg}-\mathbf{gb^\intercal}-\mathbf{s}^{-2}\mathbf{c}\mathbf{s}^{-2}+\mathbf{g}\mathbf{c}\mathbf{g}.\nonumber \\
\label{eq:lens_dot}
\end{eqnarray}
We will make use of the equations (\ref{eq:s-2_dot}) and (\ref{eq:lens_dot}) while studying the time evolution of hydrodynamic and thermodynamic variables in the following sections.

Before moving on to a hydrodynamic interpretation, we will now have a consistency check. Those results will be very useful in analysing the energy definitions presented in our work.

\subsubsection{Consistency check: a pathway to thermodynamics}\label{sec:Consistency check: a pathway to thermodynamics}
In \cite{Littlejohn:1985}, Littlejohn argues that even though Gaussian states have Gaussian Wigner functions, the converse is not necessarily true. This means that the equality of two Wigner functions does not immediately imply the equality of the corresponding wave functions. In order for a Gaussian wave function that is obtained from a Wigner distribution to match the wave function obtained via the metaplectic operators acting on a ground state wave function, one needs to introduce a phase factor. Note that this phase factor is $-\alpha/2$ which follows from the eigenvalues of the underlying fractional Fourier transformer of the classical phase space evolution that we presented in eq.~(\ref{eq:deter_AiB}). 

We will now have a consistency check which might seem redundant at a first glance. However, during this process, we will obtain the value of $\alpha$ in terms of the elements of the Hamiltonian matrix, $\mathbf{\mathbf{L}_{\mathbf{H}}}$, and the covariance matrix, $\mathbf{\Sigma}$. This result will be crucially important when we introduce the hydrodynamic interpretation and the associated thermodynamic variables in the following sections. 

It is known that in order to obtain the wave function of the squeezed coherent states, one can follow an alternative route to the one of Littlejohn. For example, one method is to start with an \textit{ansatz}, such that the wave function is in the following form 
\begin{eqnarray}\label{eq:Wavefn_Combescure}
\Phi \left(\mathbf{q},t\right)=\mathcal{A}(t)\exp{\left(\frac{i}{2\hbar}\mathbf{q}^\intercal \mathbf{\Gamma}\mathbf{q}\right)},
\end{eqnarray}
during any point of the evolution.
Here, $\mathbf{\Gamma}$ belongs to the Siegel space of complex symmetric matrices which is given as before, i.e., $\mathbf{\Gamma}=\left(\mathbf{C}+i\mathbf{D}\right)\left(\mathbf{A}+i\mathbf{B}\right)^{-1}$. Then, it is shown that in order for $\Phi$ to be a solution of the Schr{\"{o}}dinger equation, one needs to satisfy the following two conditions \cite{Combescure:2012}
\begin{eqnarray}
\frac{d\mathbf{\Gamma}}{dt}&=&-\mathbf{a}-\mathbf{\Gamma}\mathbf{b}^\intercal-\mathbf{b}\mathbf{\Gamma}-\mathbf{\Gamma}\mathbf{c}\mathbf{\Gamma},\label{eq:Gamma_dot_Combescure}\\
\frac{d\mathcal{A}}{dt}&=&-\frac{1}{2}{\rm{Tr}}\left(\mathbf{b}+\mathbf{c}\mathbf{\Gamma}\right)\mathcal{A},\label{eq:A_dot_Combescure}
\end{eqnarray}
where $\mathcal{A}$ is given as $\mathcal{A}=\left(\pi \hbar\right)^{(-n/4)}\left({\rm{det}\left[\mathbf{A}+i\mathbf{B}\right]}\right)^{-1/2}$. Here, ``${\rm{Tr}}$'' represents the trace operator. Then, the condition (\ref{eq:Gamma_dot_Combescure}) of \cite{Combescure:2012} corresponds to our eqs.~(\ref{eq:s-2_dot}) and (\ref{eq:lens_dot}) due to $\mathbf{\Gamma}$ being decomposed into its pure real and pure imaginary components in eq.~(\ref{eq:Gamma_lens_magn}). Likewise, as we obtained $\sqrt{{\rm{det}}\left(\mathbf{A}+i\mathbf{B}\right)}=\exp{\left(i\alpha/2\right)}\sqrt{{\rm{det}}\,\mathbf{s}}$ previously in eq.~(\ref{eq:deter_AiB}), the condition (\ref{eq:A_dot_Combescure}) implies
\begin{eqnarray}
\frac{d\alpha (t)}{dt}={\rm{Tr}}\left(\mathbf{c}\mathbf{s}^{-2}\right).
\end{eqnarray}
We will elaborate on the importance of this result in Section~(\ref{sec:Relationship between the internal energy and the Maslov index}).

\subsection{Probability distributions}\label{sec:Probability distributions and entropies}
\subsubsection{Phase space distribution and the entropy}\label{sec:Phase space distribution and the entropy}
Whether it is within the classical or within the quantum theory, there is no unique way of approaching the concept of probability distributions and the entropy in general. For instance, as it is stressed many times in the literature, coherent states are the minimum uncertainty states. Thus, they contain maximum information and minimum entropy. This statement is usually vaguely stated in the literature without specifying in which manner the concept of information and entropy are defined. Here, we suggest certain definitions by considering the phase space as our main object. 

From now on, we will adopt the nomenclature and the notation of statistical mechanics. For example, for a Gaussian probability distribution,
\begin{eqnarray}\label{eq:Gaussian}
\rho=\frac{1}{\left(2\pi\right)^{d/2}}\frac{1}{\sqrt{{\rm{det}}\mathbf{{M}}}}\exp{\left[-\frac{1}{2}\left(\mathbf{x}-\boldsymbol{\mu}\right)^\intercal\mathbf{{M}}^{-1}\left(\mathbf{x}-\boldsymbol{\mu}\right)\right]},
\end{eqnarray}
we will write
\begin{eqnarray}\label{eq:Gaussian_dist_stat}
\rho:=\mathcal{N}\left(\mathbf{x|\boldsymbol{\mu}},\mathbf{M}\right),
\end{eqnarray}
where $\mathbf{x}$ is a $d$--dimensional variable vector and $\boldsymbol{\mu}$ is its mean with respect to the Gaussian distribution, (\ref{eq:Gaussian}). The $d\times d$ matrix $\mathbf{M}$ is a positive semi--definite covariance matrix of the distribution. We will refer to our Gaussian Wigner function as a \textit{joint probability distribution}. This is due to it including the information about both $\mathbf{q}$ and $\mathbf{p}$ subspaces. We will denote this joint distribution associated with the squeezed coherent states, i.e., eq.~(\ref{eq:Wignerfnc}), in the short form
\begin{eqnarray}\label{eq:joint_distribution}
\rho_{(\mathbf{q},\mathbf{p})}:=W\left(\mathbf{z}\right)=\mathcal{N}\left(\mathbf{z|\langle \mathbf{\hat{z}}\rangle},\boldsymbol{\Sigma}\right),
\end{eqnarray}
where $2n$ dimensional $\langle \mathbf{\hat{z}}\rangle$ are the standard expectation values obtained by the averaging through the entire phase space, i.e., eq.~(\ref{eq:ave_z}). The covariance matrix $\boldsymbol{\Sigma}$ is given in its explicit form in eq.~(\ref{eq:Covariance_matrix_coh}) for a squeezed coherent state. Note that the phase space probability distribution, $\rho_{(\mathbf{q},\mathbf{p})}$, is normalized.

We will now define a macroscopic \textit{joint entropy}, $\mathbb{S}_{(\mathbf{q},\mathbf{p})}$, of the system which we write as
\begin{eqnarray}\label{eq:joint_entropy}
\mathbb{S}_{(\mathbf{q},\mathbf{p})} = -k_B\int \rho_{(\mathbf{q},\mathbf{p})} \ln{\rho_{(\mathbf{q},\mathbf{p})}}d\mathbf{q}d\mathbf{p},
\end{eqnarray}
by considering $\rho_{(\mathbf{q},\mathbf{p})}$ as in eq.~(\ref{eq:joint_distribution}).
Computation of the integral in eq.~(\ref{eq:joint_entropy}) now gives
\begin{eqnarray}\label{eq:Joint_ent_calc}
\mathbb{S}_{(\mathbf{q},\mathbf{p})}&=&\frac{k_B}{2}\ln{\left({\rm{det}}\left[2\pi e\frac{\hbar}{2}\mathbf{W}^{-1}\right]\right)}
=k_Bn\left(1+\ln\right[\pi \hbar \left]\right)+\frac{k_B}{2}
\underbrace{\ln \left({\rm{det}}\left[\mathbf{W}^{-1}\right]\right)}_{=0},
\end{eqnarray}
which corresponds to an equilibrium entropy as we have ${d\mathbb{S}_{(\mathbf{q},\mathbf{p})}}/{dt}=0.$
The second term on the right hand side of  eq.~(\ref{eq:Joint_ent_calc}) vanishes due to the Wigner matrix being a symplectic matrix (and so is its inverse). 
Note that $\mathbb{S}_{(\mathbf{q},\mathbf{p})}$ is indeed the Shannon entropy of the Wigner function (multiplied by $k_B$)\footnote{Information entropy matches the thermodynamic entropy up to the factor $k_B$ for systems in equilibrium as in our case.} which is sometimes referred to as the Wigner entropy \cite{Santos:2017, Brunelli:2018, Malouf:2019, Belenchia:2020}. It matches the R{\'e}nyi$-2$ entropy up to a constant for Gaussian states \cite{Adesso:2012}. It also corresponds to the lower bound of the missing position and momentum information for a stationary system as presented in \cite{Birula:1975} and whose relation to the Heisenberg uncertainty principle has been discussed in \cite{Birula:2011}.

Indeed, the fact that $\mathbb{S}_{(\mathbf{q},\mathbf{p})}$ corresponds to a minimum entropy state can be argued within the Schr{\"{o}}dinger--Robinson uncertainty principle \cite{deGosson:2009, deGosson:2019}. Namely, the determinant of the covariance matrix, i.e.,
\begin{eqnarray}
{\rm{det}}\boldsymbol{\Sigma}&=&{\rm{det}}\left(\boldsymbol{\sigma_{qq}}\right){\rm{det}}\left(\boldsymbol{\sigma_{pp}}-\boldsymbol{\sigma_{pq}}\boldsymbol{\sigma_{qq}}^{-1}\boldsymbol{\sigma_{qp}}\right)
=\left(\frac{\hbar}{2}\right)^{2n}
\end{eqnarray}
corresponds to the minimum of the Schr{\"{o}}dinger--Robinson uncertainty 
\begin{eqnarray}
{\sigma_{qq}{\sigma_{pp}}\ge \sigma_{pq}}{\sigma_{qp}}+\frac{\hbar ^{2}}{4},
\end{eqnarray}
which was originally defined for a 1--dimensional configuration space.
Thus, phase space entropy taking its minimum value is consistent with the minimum uncertainty and maximum information accommodated by the squeezed coherent states. Note that for those states, if there exists no classical phase space shearing/lensing, i.e., $\mathbf{g}=\mathbf{0}$ in eq.~(\ref{eq:Covariance_matrix_coh}), one has $\boldsymbol{\sigma_{pq}}=\boldsymbol{\sigma_{qp}}^\intercal=\mathbf{0}$ and the minimum of the standard Heisenberg uncertainty is reached \footnote{Also see the discussions on correlated states and generalized uncertainty relations in \cite{Dodonov:1980nx} in relation to this specific case.}.

Moreover, $\mathbb{S}_{(\mathbf{q},\mathbf{p})}$ being a minimum seems to be consistent with certain topological arguments. Previously, it was recognized by de Gosson \cite{deGosson:2006, deGossonLuef:2009, deGosson:2013, deGosson:2013b} that symplectic non-squeezing theorem of Gromov \cite{Gromov:1985} can be realized to define some minimum uncertainty units on phase space. Those are known as the \textit{quantum blobs}. Namely, on the plane of conjugate canonical pairs, there exist a minimum area of size $\pi \hbar$. Due to the underlying symplectic capacity, the canonical pairs that compose the projected area can not take lower values. Here, we suggest that $\mathbb{S}_{(\mathbf{q},\mathbf{p})}$ reflects the missing information contained in the quantum blobs of de Gosson.

\subsubsection{Marginal and conditional distributions}\label{sec:Marginal and conditional distributions}
Even though phase space methods work surprisingly well, at least for the Gaussian states, it is the position space that we have immediate experimental access to. Naturally, the de Broglie--Madelung--Bohm theory was originally presented in the $\mathbf{q}$--representation. This requires the introduction of marginal and conditional objects for the statistical considerations of quantum mechanics. For instance, in \cite{Moyal:1949}, Moyal introduced \textit{space--conditional averages} for the observables.  Accordingly, Takabayashi argued that the quantum potential can be considered as an apparent agent emerging from the configuration space projections of the phase space distributions \cite{Takabayasi:1954}.

Consequently, we introduce our marginal and conditional probability distributions now. Let us write the marginal distribution as
\begin{eqnarray}
\rho_{(\mathbf{q})}=\int \rho_{(\mathbf{q},\mathbf{p})}d\mathbf{p},\label{eq:marginal_distribution}
\end{eqnarray}
where $\rho_{(\mathbf{q})}$ is the $\mathbf{q}$--space probability distribution whose value is equal to $\left[R\left(\mathbf{q},t\right)\right]^2$. The average of a function $f=f(\mathbf{q},\mathbf{p})$ over the marginal distribution is obtained by
\begin{eqnarray}\label{eq:marginal_ave}
{\langle f \rangle}_{(\mathbf{q})}=\int f \rho_{(\mathbf{q})}d{\mathbf{q}}.
\end{eqnarray}

However, $\rho_{(\mathbf{q})}$ includes only the information that is needed to describe the position coordinates. Once the positions are known, the remaining, additional information needed in order to specify $\mathbf{p}$ is obtained by a conditional distribution which is sometimes referred to as a \textit{posterior distribution}. We write the conditional probability distribution as
\begin{eqnarray}
\rho_{(\mathbf{p}|\mathbf{q})}&=&\rho_{(\mathbf{q},\mathbf{p})}/\rho_{(\mathbf{q})}\label{eq:cond_distribution}.
\end{eqnarray}
The average of a function $f=f(\mathbf{q},\mathbf{p})$ over the conditional distribution can now be obtained by
\begin{eqnarray}\label{eq:cond_ave}
{\langle f \rangle}_{(\mathbf{p}|\mathbf{q})}=\int f \rho_{(\mathbf{p}|\mathbf{q})}d{\mathbf{p}}.
\end{eqnarray}
Once we compute the values of the marginal and the conditional distributions for the squeezed coherent states, we get
\begin{eqnarray}\label{eq:marg_cond_distribution}
\rho_{(\mathbf{q})}=\mathcal{N}\left(\mathbf{q|\langle \mathbf{\hat{q}}\rangle}_{(\mathbf{q})},\frac{\hbar}{2}\mathbf{s}^{2}\right),\qquad {\rm{and}} \qquad \rho_{(\mathbf{p}|\mathbf{q})}=\mathcal{N}\left(\mathbf{p|\langle \mathbf{\hat{p}}\rangle}_{(\mathbf{p}|\mathbf{q})},\frac{\hbar}{2}\mathbf{s}^{-2}\right).
\end{eqnarray}
Here,
\begin{eqnarray}\label{eq:marg_cond_averages}
{\langle \mathbf{\hat{q}}\rangle}_{(\mathbf{q})}={\langle \mathbf{\hat{q}}\rangle}, \qquad 
{\rm{and}}
\qquad {\langle \mathbf{\hat{p}}\rangle}_{(\mathbf{p}|\mathbf{q})}=\langle \mathbf{\hat{p}}\rangle-\mathbf{g}\left(\mathbf{q}-\langle \mathbf{\hat{q}}\rangle\right)
\end{eqnarray}
are the mean values taken with respect to the marginal and the conditional distributions respectively. 
As we argued in Section~(\ref{sec:Phase space distribution and the entropy}) Schr{\"{o}}dinger--Robinson uncertainty principle is more viable for generic squeezed coherent states as for a generic case the minimum of the Heisenberg uncertainty is not satisfied. At a first glance, this seems to be contradicting with the idea of a minimum entropy state. However, Littlejohn attributes the minimum Heisenberg uncertainty not being reached by the squeezed coherent states to a geometric explanation \cite{Littlejohn:1985}. Specifically, it is due to the choice of a wrong symplectic frame. He argues that it is only when the principal axes of the Wigner ellipsoid coincide with the axes of positions and momenta, the minimum Heisenberg uncertainty is achieved. If they are not aligned, the angles of projections of the Wigner ellipsoid on the phase space planes cause the system to appear as if it is not at a minimum uncertainty state. 

Note that the original covariance matrix, $\boldsymbol{\Sigma}=\frac{\hbar}{2}\mathbf{W}^{-1}$, takes a block diagonal form
\begin{eqnarray}\label{eq:Cov_mat_diag}
\boldsymbol{\Sigma}
=
\frac{\hbar}{2}
\left[
\begin{array}{c|c}
\mathbf{s^2} & \, \, \mathbf{0} \\
\hline
\mathbf{0} & \, \, \mathbf{s}^{-2}
\end{array}
\right],
\end{eqnarray}
when the lensing/shearing matrix satisfies $\mathbf{g}=\mathbf{0}$. This is when ${\langle\mathbf{\hat{p}}\rangle}={\langle\mathbf{\hat{p}}\rangle}_{(\mathbf{p}|\mathbf{q})}$. Thus, our conditional distribution $\rho_{(\mathbf{p}|\mathbf{q})}$, which has a variance matrix $\frac{\hbar}{2}\mathbf{s}^{-2}$ for all cases, keeps track of the minimum Heisenberg uncertainty with respect to conditional momenta.

Before going further into the thermodynamic interpretation, we will now investigate more on the probability distributions and their time evolutions.
\subsubsection{The Fokker--Planck equation, probability fluxes and the continuity equation}\label{sec:Fokker-Planck equation, probability fluxes and the continuity equation}
In statistical mechanics, the Fokker--Planck equation is considered as a stochastic differential equation that gives the time evolution of a probability distribution, $\rho\left(\mathbf{x},t\right)$, and which follows as
\begin{eqnarray}
\frac{\partial \rho}{\partial t}=-{\nabla}{_{x_i}}\left(\boldsymbol{\beta }_i\rho\right)+{\nabla}{_{x_i}}{\nabla}{_{x_j}}\left(\mathbfcal{D}_{ij}\rho \right),
\end{eqnarray}
where $\boldsymbol{\beta}\left(\mathbf{x},t\right)$ is the drift vector and $\mathbfcal{D}\left(\mathbf{x},t\right)$ is the diffusion matrix.
It is known that Gaussian distributions are exact solutions of the Fokker--Planck equations. For a generic multi--dimensional Gaussian distribution, 
$\rho=\mathcal{N}\left(\mathbf{x|\boldsymbol{\mu}},\mathbf{M}\right)$ with $\mathbf{M}=\mathbf{M}(t)$ as in eq.~(\ref{eq:Gaussian_dist_stat}), the corresponding drift vector and the diffusion matrix  are
\begin{eqnarray}\label{eq:Drift_Diffusion}
\boldsymbol{\beta}(t)=\frac{d\boldsymbol{\mu}}{dt}, \qquad \qquad \mathbfcal{D}(t)=\frac{1}{2}\frac{d\mathbf{M}}{dt}.
\end{eqnarray}
such that the Fokker--Planck equation can be written as
\begin{align}\label{eq:Fokker-Planck-multidimen}
\frac{\partial \rho}{\partial t}=\left[ 
\frac{1}{2}\left(\mathbf{x}-\boldsymbol{\mu}\right)^\intercal\mathbf{M}^{-\intercal}\frac{d\mathbf{M}}{dt}\mathbf{M}^{-1}\left(\mathbf{x}-\boldsymbol{\mu}\right)+\frac{d\boldsymbol{\mu}}{dt}^\intercal\mathbf{M}^{-1}\left(\mathbf{x}-\boldsymbol{\mu}\right)-\frac{1}{2}{\rm{Tr}}\left(\frac{d\mathbf{M}}{dt}\mathbf{M}^{-1}\right)\right]\rho.
\end{align}
Equation~(\ref{eq:Fokker-Planck-multidimen}) holds for our joint distribution, $\rho_{(\mathbf{q},\mathbf{p})}=\mathcal{N}\left(\mathbf{z}|\langle \mathbf{\hat{z}}\rangle,\boldsymbol{\Sigma}\right)$, our marginal distribution $\rho_{(\mathbf{q})}=\mathcal{N}\left(\mathbf{q|\langle \mathbf{\hat{q}}\rangle}_{(\mathbf{q})},\frac{\hbar}{2}\mathbf{s}^{2}\right)$, and our conditional distribution, $
\rho_{(\mathbf{p}|\mathbf{q})}=\mathcal{N}\left(\mathbf{p|\langle \mathbf{\hat{p}}\rangle}_{(\mathbf{p}|\mathbf{q})},\frac{\hbar}{2}\mathbf{s}^{-2}\right)$ once we replace the variable vector, the mean vector and the covariance matrix in eq.~(\ref{eq:Fokker-Planck-multidimen}) with the desired ones.

Let us now recall from the brief summary of the de Broglie--Madelung--Bohm approach given in Section~(\ref{sec:de Broglie-Bohm approach}) that the pure imaginary part of the Schr{\"{o}}dinger equation can be interpreted as a continuity equation, (\ref{eq:cont_deBroglie-Madelung-Bohm}), given that the squared amplitude of the wave function, $\left[R\left(\mathbf{q},t\right)\right]^2$,
 is interpreted as the density of a fluid. Within the statistical interpretation, it gives the \textit{the} probability amplitude of a given outcome. Thus, it is no surprise that the Fokker--Planck equation, (\ref{eq:Fokker-Planck-multidimen}), of our marginal distribution, $\rho_{(\mathbf{q})}$, is in the same footing as the imaginary part of the Schr{\"{o}}dinger equation given in eq.~(\ref{eq:Sch_cmplx}), multiplied by $R\left(\mathbf{q},t\right)$. 

On the other hand, every probability distribution that satisfies a Fokker--Planck equation has an associated probability flux that satisfies a continuity equation. What we want to investigate here is to see whether the Fokker--Planck induced flux term is the same as the flux term that appears in the hydrodynamic interpretation of the Schr{\"{o}}dinger equation. 

Namely, for a generic Gaussian $\rho=\mathcal{N}\left(\mathbf{x|\boldsymbol{\mu}},\mathbf{M}\right)$ which satisfies the Fokker--Planck equation, (\ref{eq:Fokker-Planck-multidimen}), one can define a probability flux
\begin{eqnarray}
\mathbf{j}=\boldsymbol{\beta}\rho-\mathbfcal{D}\boldsymbol{\nabla _x}\rho,
\end{eqnarray}
with the drift vector and the diffusion matrix given in eq.~(\ref{eq:Drift_Diffusion}) such that an associated continuity equation
\begin{eqnarray}
\frac{\partial \rho}{\partial t}+\boldsymbol{\nabla _x}^\intercal\mathbf{j}=0
\end{eqnarray}
is satisfied.

For our marginal distribution, for example, the continuity equation
\begin{eqnarray}\label{eq:marginal_Fokker-Planck}
\frac{\partial \rho_{(\mathbf{q})}}{\partial t}+\boldsymbol{\nabla}_{\mathbf{q}}^\intercal\mathbf{j}_{(\mathbf{q})}=0
\end{eqnarray}
is a Fokker--Planck equation with an associated flux
\begin{eqnarray}\label{eq:marginal_flux}
\mathbf{j}_{(\mathbf{q})}&=&\frac{d\langle \mathbf{\hat{q}}\rangle}{dt}\rho_{(\mathbf{q})}-\frac{\hbar}{4}\frac{d\left(\mathbf{s}^2\right)}{dt}\boldsymbol{\nabla}_{\mathbf{q}} \rho_{(\mathbf{q})}
=
\left[\mathbf{b}\langle \mathbf{\hat{q}}\rangle+\mathbf{c}\langle \mathbf{\hat{p}}\rangle+\frac{1}{2}\frac{d\left(\mathbf{s}^2\right)}{dt}\mathbf{s}^{-2}\left(\mathbf{q}-\langle \mathbf{\hat{q}}\rangle\right)\right]\rho_{(\mathbf{q})}.
\end{eqnarray}
This result follows from (i) the phase space expectation values following classical trajectories, i.e., ${d\langle \mathbf{\hat{z}}\rangle}/{dt}={\rm{\mathbf{L_H}}}\langle \mathbf{\hat{z}}\rangle$, (ii) $\boldsymbol{\nabla}_{\mathbf{q}} \rho_{(\mathbf{q})}=-2\hbar^{-1}\mathbf{s}^{-2}\left(\mathbf{q}-\langle \mathbf{\hat{q}}\rangle\right)\rho_{(\mathbf{q})}$ being satisfied with $\rho_{(\mathbf{q})}$ given in  eq~(\ref{eq:marg_cond_distribution}), and (ii)
\begin{eqnarray}\label{eq:ds^2dts^-2}
\frac{d\left(\mathbf{s}^2\right)}{dt}\mathbf{s}^{-2}=\left(\mathbf{b}^\intercal\mathbf{s}^2+\mathbf{s}^2\mathbf{b}-\mathbf{cgs}^2-\mathbf{s}^2\mathbf{gc}\right)\mathbf{s}^{-2},
\end{eqnarray}
via the evolution of the phase space covariance matrix, eq.~(\ref{eq:Sigma_dot}), as indicated before.

Let us now compare the flux, $\mathbf{j}_{(\mathbf{q})}$, in eq.~(\ref{eq:marginal_flux}) with the one of the de Broglie--Madelung--Bohm theory, $\mathbf{\tilde{j}}$, introduced in Section~(\ref{sec:de Broglie-Bohm approach}).
Note that the latter is defined explicitly for a specific Hamiltonian operator as in the Schr{\"{o}}dinger equation (\ref{eq:psi_schr_causal}). Moreover, as discussed in \cite{Deotto:1997}, its definition given by eq.~(\ref{eq:flux_deBroglie-Madelung-Bohm}) is non--unique. Also as discussed in \cite{Wiseman:2007} there exists an arbitrariness on the definition of a probability current in general. 
In order to make a connection with the generic Hamiltonian here, we will take $1/m\rightarrow \mathbf{c}$ where $m$ refers to the mass of the particle in the standard approach, as the matrix $\mathbf{c}$ is responsible for the coupling of the momentum operator in our Hamiltonian in eq.~(\ref{eq:Ham_gen}). Also, as  $\left[R\left(\mathbf{q},t\right)\right]^2=\rho_{(\mathbf{q})}$  we have $\mathbf{\tilde{j}}=\rho_{(\mathbf{q})}\mathbf{\tilde{v}}=\rho_{(\mathbf{q})}\mathbf{c}\mathbf{\tilde{p}}$ where $\mathbf{\tilde{v}}$ is considered as the linear velocity. The term $\mathbf{\tilde{p}}=\boldsymbol{\nabla}_{\mathbf{q}} \mathcal{S}$ are sometimes referred to as the Bohm momenta and they are expected to satisfy the equation of motion (\ref{eq:eom_deBroglie-Madelung-Bohm}). However, we emphasise that the original de Broglie--Madelung--Bohm theory was constructed with such a choice of Hamiltonian operator that the resultant Schr{\"{o}}dinger equation is interpreted within a hydrodynamic interpretation of an \textit{irrotational} fluid flow. That is why the momenta $\mathbf{\tilde{p}}$ can be written as a divergence of a potential/phase. We will discuss the actual meaning of $\mathbf{\tilde{p}}$ in a short while. We should first emphasize that our generic Hamiltonian operator, (\ref{eq:Ham_gen}), includes some position--momentum coupling terms which are associated with the rotational degrees of freedom in general. Therefore, the probability flux for the marginal distribution, $\mathbf{j}_{(\mathbf{q})}$, includes the fluxes associated with the rotations in addition to those associated with the linear motion.

In order to show this explicitly, let us assume in eq.~(\ref{eq:ds^2dts^-2}), the symmetry of the products $\mathbf{b}^\intercal\mathbf{s}^2$ and $\mathbf{cgs}^2$ for our symmetric matrices $\mathbf{s}$, $\mathbf{c}$ and $\mathbf{g}$. Then, the flux associated with the marginal distribution can be written as
\begin{eqnarray}
\mathbf{j}_{(\mathbf{q})}=\mathbf{j}^{\rm{irrot.}}_{(\mathbf{q})}+\mathbf{j}^{\rm{rot.}}_{(\mathbf{q})},
\end{eqnarray}
with
\begin{eqnarray}
\mathbf{j}^{\rm{irrot.}}_{(\mathbf{q})}=\rho_{(\mathbf{q})}\mathbf{c}{\langle \mathbf{\hat{p}}\rangle}_{(\mathbf{p}|\mathbf{q})},
\qquad
{\rm{and}}
\qquad
\mathbf{j}^{\rm{rot.}}_{(\mathbf{q})}=\rho_{(\mathbf{q})}\left[\mathbf{b}{\langle \mathbf{\hat{q}}\rangle}+\mathbf{b}^\intercal  \left(\mathbf{q}-{\langle \mathbf{\hat{q}}\rangle}\right)\right].
\end{eqnarray}
Recall that the position--momentum coupling is represented by the matrix $\mathbf{b}$ in the generic Hamiltonian. Therefore, $\mathbf{j}^{\rm{rot.}}_{(\mathbf{q})}$ is the portion of the marginal probability flux that includes only those coupling terms. In the case that $\mathbf{b}$ is symmetric, a velocity term can be associated with rotational degrees of freedom, $\mathbf{v}_{\rm{rot.}}=\mathbf{j}^{\rm{rot.}}_{(\mathbf{q})}/\rho_{(\mathbf{q})}=\mathbf{b}\mathbf{q}$ which takes a local form. This is similar to a tangential velocity field for a rotational flow. Obviously, when $\mathbf{b}=0$, it is only the irrotational velocity $\mathbf{v}_{\rm{irrot.}}=\mathbf{j}^{\rm{irrot.}}_{(\mathbf{q})}/\rho_{(\mathbf{q})}=\mathbf{c}{\langle \mathbf{\hat{p}}\rangle}_{(\mathbf{p}|\mathbf{q})}$ that governs the dynamics of the ensemble. Here, ${\langle \mathbf{\hat{p}}\rangle}_{(\mathbf{p}|\mathbf{q})}=\langle \mathbf{\hat{p}}\rangle-\mathbf{g}\left(\mathbf{q}-\langle \mathbf{\hat{q}}\rangle\right)$ are the conditionally averaged momenta given by eq.~(\ref{eq:marg_cond_averages}) previously. It can easily be checked that ${\langle\mathbf{\hat{p}}\rangle}_{(\mathbf{p}|\mathbf{q})}=\boldsymbol{\nabla}_{\mathbf{q}}\mathcal{S}$ holds by making use of the phase function, $\mathcal{S}$, eq.~(\ref{eq:phase_sq_coh}), of the squeezed coherent state wave function. This means that the conditionally averaged momenta are equal to the Bohm's momenta, $\mathbf{\tilde{p}}$, and $\mathbf{j}^{\rm{irrot.}}_{(\mathbf{q})}$ are equivalent to $\mathbf{\tilde{j}}$ as expected. 

It is known that within a weak measurement of Aharonov \textit{et al.} \cite{Aharonov:1988}, the post--selection of positions does not completely destroy the momentum information and a mean, weak value of momentum can be obtained for a system of particles. A weak value is in general a complex number and its real part is obtained by averaging the desired observable conditioned on a second measurement \cite{Dressel:2010,Dressel:2012,Dressel:2012b}. In this work, we consider momentum conditionally averaged on positions following Moyal's \cite{Moyal:1949} and Sonego's \cite{Sonego:1991} arguments.
Also, it was previously realized by many researchers that $\mathbf{\tilde{p}}$ are the real, measurable part of the weak value of the momenta \cite{Hiley:2012w,Feyereisen:2015,Flack:2018}. Therefore, the conditionally averaged momenta act as an \textit{effective} momenta of a system. It is known that one can imagine a flow of an ensemble of particles such that ${\langle\mathbf{\hat{p}}\rangle}_{(\mathbf{p}|\mathbf{q})}=\mathbf{\tilde{p}}$ represent a stream--line momenta rather than the momenta of the individual particles \cite{Sonego:1991}. Our last remark is that it is the matrix $\mathbf{g}$, that is responsible for the classical shearing/lensing effect, which differentiates the results of strong and weak measurements of momenta for squeezed coherent states. 

Let us now compare our set up with some investigations in the literature that discuss stochastic quantum mechanics in relation to Einstein's work on Brownian motion \cite{Einstein:1905, Einstein:1956, Einstein:1990}. The important point we would like to start highlighting here is that Einstein used some simplifying assumptions while building up his theory. For simplicity, he assumed the suspension of a particle within a homogeneous and \textit{stationary} liquid. Moreover, the time scale, over which the dynamics takes place is assumed to be smaller than the observation time. In addition, random displacements of the particle is assumed to be small in order to give a time independent probability distribution for the displacements. \textit{Only then}, his Fokker--Planck equation, which is also known as the diffusion equation, takes its simple form,
\begin{eqnarray}\label{eq:Diff_eq_Einstein_Diffu}
\frac{\partial f}{\partial t}=D\nabla ^2f.
\end{eqnarray}
Here, the distribution function $f=f(q,t)$  reflects the number of particles per unit volume in 1-dimension, $\bm{\nabla}$ is taken with respect to the position, and $D$ is a \textit{constant} diffusion coefficient, contrary to our time dependent diffusion matrices presented before. 
Then, the solution of eq.~(\ref{eq:Diff_eq_Einstein_Diffu}) for the distribution function is given by
\begin{eqnarray}
f(q,t)=\frac{1}{\sqrt{4\pi D t}}\exp{\left(-\frac{q^2}{4Dt}\right)},
\end{eqnarray}
with $\langle q^2\rangle=2Dt$ being the second moment of the displacements. Next, a drift effect can be added by hand if there exists a constant external force acting on the particle in order to balance the diffusion effect. This balance can be written as
\begin{eqnarray}\label{eq:flux_balance}
\mathbf{j}=\mathbf{j}_{\rm{drift}}+\mathbf{j}_{\rm{diffusion}}=0\qquad \rightarrow \qquad \bm{\beta} f=D{\bm{\nabla}} f
\end{eqnarray}
such that
\begin{eqnarray}
\frac{\partial f}{\partial t}+{\bm{\nabla}}\mathbf{j}\rightarrow \frac{\partial f}{\partial t}=0.
\end{eqnarray}
In their formulation of quantum stochastic theory, Bohm and Hiley define a generic theory which, in its equilibrium, corresponds to standard quantum mechanics \cite{Bohm:1989,Bohm:1995}. This is achieved by adding a stochastic contribution by hand to the standard probability flux $\mathbf{\tilde{j}}$ of the de Broglie--Madelung--Bohm theory. In order to form an analogous theory to Brownian motion, they refer to its version in Einstein's work in which a particle is suspended in liquid under the gravitational force \cite{Einstein:2006}. The osmotic/drift velocity and the diffusion flux they introduce are in the same form as they appear in Einstein's original formulation. However, recall that Einstein's work was constructed on a system restricted by some assumptions \footnote{In fact, those assumptions result in certain mathematical inconsistencies, including the breakdown of Galilean invariance as discussed in \cite{Ryskin:1997}.}. Though, in Bohm and Hiley's work, the system does not necessarily have to fall under such category.

Now, let us consider our marginal probability flux, $\mathbf{j}_{(\mathbf{q})}$, given in eq.~(\ref{eq:marginal_flux}) which satisfies a more generic Fokker--Plack equation than the one of Einstein. An analogous balance equation as in eq.~(\ref{eq:flux_balance}) is satisfied without introducing an extra flux term when
\begin{eqnarray}
\mathbf{c}\left(\langle \mathbf{\hat{p}}\rangle-\mathbf{g}\left(\mathbf{q}-\langle \mathbf{\hat{q}}\rangle\right)\right)=\mathbf{c}{\langle \mathbf{\hat{p}}\rangle}_{(\mathbf{p}|\mathbf{q})}=\mathbf{c}\boldsymbol{\nabla}_{\mathbf{q}}\mathcal{S}=0,
\end{eqnarray}
if $\mathbf{b}=0$ and $\mathbf{cgs}^2$ is symmetric. This refers to a stationary state in which the system of particles has zero average velocity. Thermodynamically speaking, this should be reached at some absolute zero temperature. Note that such a result is consistent with the quantum equilibrium condition defined in  \cite{Bohm:1989,Bohm:1995} and it is obtained without using the simplifying assumptions of Einstein. Thus, it is a curious subject whether a more generic stochastic theory can be developed by considering relatively more generic Fokker--Planck equations as presented here. 

The last point we would like to emphasise is that even though the Fokker--Planck equation is usually associated with stochastic processes, it would be misleading to interpret the evolution of the probability distributions of the generalized squeezed coherent states investigated here in a stochastic manner. Note that the drift vectors of the Gaussian distributions here are obtained through the time evolution of the averages of the corresponding phase space vectors. Those expectation values follow the same path as the classical trajectories in phase space. Moreover, the associated diffusion matrices of our distributions are obtained via the time evolution of the corresponding covariance matrices whose relation to the classical phase space magnifications and the classical shears has been established in the previous sections. As the diffusion matrices of neither the joint, the marginal, nor the conditional distributions of ours can be considered as random matrices, the evolution of the corresponding probability distributions can not simply be interpreted as a stochastic process.

\subsection{Thermodynamic variables}\label{sec:Thermodynamic variables}
\subsubsection{Back to entropy: Sackur \& Tetrode}\label{sec:An analogue Sackur–Tetrode entropy}
Previously, in Section~(\ref{sec:Phase space distribution and the entropy}), we defined a joint phase space distribution, $\rho_{(\mathbf{q},\mathbf{p})}$, and an equilibrium entropy, $\mathbb{S}_{(\mathbf{q},\mathbf{p})}$. Later, in Section~(\ref{sec:Marginal and conditional distributions}) we defined a marginal distribution, $\rho_{(\mathbf{q})}$, for positions and a conditional distribution, $\rho_{(\mathbf{p}|\mathbf{q})}$, for momenta. Now we will define their associated entropies regarding an analogy between the classical Sackur–Tetrode entropy \cite{Sackur:1911,Tetrode:1912} of the kinetic theory, $S^{\rm{ST}}$,  and $\mathbb{S}_{(\mathbf{q},\mathbf{p})}$.

For a system of particles, which is represented by the Boltzmann statistics, one can identify the contributions to $S^{\rm{ST}}$ regarding the missing information of the particles' continuous positions and momenta as \footnote{See, for example, the detailed discussions given in Section 4.3, Section 5.4 and Appendix L of \cite{Ben:2008}.}
\begin{eqnarray}\label{eq:S_ST_posit_mom}
S^{\rm{ST}}_{\left({\rm{positions}}\right)}=k_BN\ln{V}=\frac{k_Bn}{2}\ln{L^2}, \qquad {\rm{and}} \qquad S^{\rm{ST}}_{\left({\rm{momenta}}\right)}=\frac{k_Bn}{2}\ln{\left(2\pi e m T\right)}.
\end{eqnarray}
Here $n=3N$ for a number of $N$ particles in $3-$dimensions. The cubical box that encloses the particles has a volume $V=L^3$. The mass of each particle is given by $m$ and $T$ is the temperature as it appears in the Boltzmann distribution. 

One then includes certain corrections to the classical $S^{\rm{ST}}$. For example,  by (i) considering a finite, discretized space obtained via dividing the continuous space into $\left(\pi \hbar\right)$ sized boxes and neglecting the contribution, $S^{\rm{ST}}_{\left({\rm{quantum}}\right)}=k_Bn\ln{\left(\pi \hbar\right)}$, coming from the quantum mechanical uncertainty within each quantum sized box, and (ii) subtracting the extra information in $S^{\rm{ST}}_{\left({\rm{positions}}\right)}$ due to assuming that the particles are distinguishable. 
Note that our phase space entropy, $\mathbb{S}_{(\mathbf{q},\mathbf{p})}$, is equal to $S^{\rm{ST}}_{\left({\rm{quantum}}\right)}$ up to a constant addition term. Now we would like to decompose $\mathbb{S}_{(\mathbf{q},\mathbf{p})}$ in a fashion similar to the decomposition of the classical part of the $S^{ST}$, i.e.,
\begin{eqnarray}
S^{\rm{ST}}_{cl.}=S^{\rm{ST}}_{\left({\rm{positions}}\right)}+S^{\rm{ST}}_{\left({\rm{momenta}}\right)}\qquad
\Longleftrightarrow 
\qquad
\mathbb{S}_{(\mathbf{q},\mathbf{p})}=\mathbb{S}_{(\mathbf{q})}+\mathbb{S}_{(\mathbf{p}|\mathbf{q})}.
\end{eqnarray}
For this, we define certain quantum entropies such that
the entropy defined through the marginal distribution, $\rho_{(\mathbf{q})}=\mathcal{N}\left(\mathbf{q|\langle \mathbf{\hat{q}}\rangle}_{(\mathbf{q})},\frac{\hbar}{2}\mathbf{s}^{2}\right)$, corresponds to the missing information due to positions. We write it as
\begin{eqnarray}\label{eq:marginal_entropy}
\mathbb{S}_{(\mathbf{q})}=-k_B\int \rho_{(\mathbf{q})}\ln{\rho_{(\mathbf{q})}}d\mathbf{q}.
\end{eqnarray}
The entropy defined through our conditional distribution, $
\rho_{(\mathbf{p}|\mathbf{q})}=\mathcal{N}\left(\mathbf{p|\langle \mathbf{\hat{p}}\rangle}_{(\mathbf{p}|\mathbf{q})},\frac{\hbar}{2}\mathbf{s}^{-2}\right)$ corresponds to the missing information due to the post--selected momenta and we write it as
\begin{eqnarray}\label{eq:cond_entropy}
\mathbb{S}_{(\mathbf{p}|\mathbf{q})}=-k_B\int \rho_{(\mathbf{q},\mathbf{p})}\ln{\rho_{(\mathbf{p}|\mathbf{q})}}d\mathbf{q}d\mathbf{p}.
\end{eqnarray}
Then, we have
\begin{eqnarray}
\mathbb{S}_{(\mathbf{q})}=\frac{k_B}{2}\ln \left({\rm{det}}\left[2\pi e \frac{\hbar}{2}\mathbf{s}^2\right]\right),\label{eq:marg_ent_calc}
\qquad
{\rm{and}}
\qquad
\mathbb{S}_{(\mathbf{p}|\mathbf{q})}=\frac{k_B}{2}\ln \left({\rm{det}}\left[2\pi e \frac{\hbar}{2}\mathbf{s}^{-2}\right]\right),\label{cond_ent_calc}
\end{eqnarray}
hence
\begin{eqnarray}
\mathbb{S}_{(\mathbf{q},\mathbf{p})}&=&\mathbb{S}_{(\mathbf{q})}+\mathbb{S}_{(\mathbf{p}|\mathbf{q})}=k_Bn\left(1+\ln\right[\pi \hbar \left]\right).\nonumber \\
\end{eqnarray}
Comparison of eqs.~(\ref{eq:S_ST_posit_mom}) and eqs.~(\ref{eq:marginal_entropy})-(\ref{eq:cond_entropy}) shows that our marginal and conditional entropies are the \textit{analogues} of $S^{\rm{ST}}_{\left({\rm{positions}}\right)}$ and $S^{\rm{ST}}_{\left({\rm{momenta}}\right)}$ within the quantum realm. The only difference is that we treat positions and momenta on an equal footing. For instance, the definition of  $S^{\rm{ST}}_{\left({\rm{momenta}}\right)}$ acknowledges the fact that there exists a variance in momenta due to the difference in the bulk motion of a system and the classical peculiar velocities of the particles. This is represented by the temperature term that appears in $S^{\rm{ST}}_{\left({\rm{momenta}}\right)}$. This is no different for our $\mathbb{S}_{(\mathbf{p}|\mathbf{q})}$ which represents the missing information due to the variance, $\frac{\hbar}{2}\mathbf{s}^{-2}$, of the conditional momenta. On the other hand, for the first case, since the system is composed of classical particles there is no spreading information within $S^{\rm{ST}}_{\left({\rm{positions}}\right)}$. Whereas, $\mathbb{S}_{(\mathbf{q})}$ is defined via the quantum mechanical position variance $\frac{\hbar}{2}\mathbf{s}^{2}$ as in the entropy for momenta.

The last point we would like to emphasise is that whether it is a classical or a quantum definition, thermodynamic entropy is a macroscopic object defined through the entire phase space. If we want to talk about a thermodynamic equilibrium, it is the joint phase space entropy that defines an equilibrium state with ${d\mathbb{S}_{(\mathbf{q},\mathbf{p})}}/{dt}=0$. However, one can expect entropy production associated with the marginal and the conditional entropies. In fact, we have
\begin{eqnarray}
\frac{d\mathbb{S}_{(\mathbf{q})}}{dt}=k_B{\rm{Tr}}\left(\mathbf{b}-\mathbf{gc}\right)=-\frac{d\mathbb{S}_{(\mathbf{p}|\mathbf{q})}}{dt},
\end{eqnarray}
due to eq.~(\ref{eq:s-2_dot}). This means that the information gain/loss in positions and in momenta are equal in magnitude and opposite in sign. They cancel each other in order for the system to satisfy a dynamic thermodynamic equilibrium at all times. 

\subsubsection{Quantum pressure and quantum temperature}\label{sec:Quantum pressure and temperature}
Now the question is whether or not the entropies we defined in the previous sections really fit into a thermodynamic picture within some analogue quantum kinetic theory. In order to investigate this, we refer to Sonego's work \cite{Sonego:1991}, in which he  presents a detailed investigation of the hydrodynamic interpretation of quantum mechanics for generic states. 

Sonego considers the standard Hamiltonian as in the original de Broglie--Madelung--Bohm method. He starts his investigation by defining a pressure tensor that is written as
\begin{eqnarray}\label{eq:Pressure_tensor}
\mathbfcal{P}=-\frac{\hbar^2}{4}\rho_{(\mathbf{q})}\mathbf{c}\boldsymbol{\nabla}_{\mathbf{q}}\boldsymbol{\nabla}_{\mathbf{q}}^\intercal \rho_{(\mathbf{q})}
\end{eqnarray}
in our notation. By adopting some techniques from the kinetic theory and by making use of the Wigner function of the phase space, he shows that the definition of the pressure tensor in eq.~(\ref{eq:Pressure_tensor}), indeed follows from a term
\begin{eqnarray}\label{eq:Temperature}
\mathbfcal{T}\left(\mathbf{q},t\right)=\frac{1}{k_B}\mathbf{c}\int \left(\mathbf{p}-{\langle\mathbf{\hat{p}}\rangle}_{(\mathbf{p}|\mathbf{q})}\right)^2\rho_{(\mathbf{p}|\mathbf{q})}d\mathbf{p}.
\end{eqnarray}
In that case, an equation of state $\mathbb{P}={\rm{Tr}}\mathbfcal{P}=\rho_{(\mathbf{q})}k_B\mathbb{T}$ is satisfied with the temperature term $\mathbb{T}={\rm{Tr}}\mathbfcal{T}$. The integral in eq.~(\ref{eq:Temperature}) essentially gives the variance of momenta via which a temperature term is defined. This is similar to the case of the classical kinetic theory. What is important here is that it is the variance of the conditionally averaged momentum that defines a macroscopic phenomenon like temperature here. This means weak measurements are again at the center of the definitions of measurable thermodynamic variables \footnote{Also see the discussion of \cite{Feyereisen:2015} on the variance of the conditional momentum and its relation to weak measurements in the context of Sonego's work.}.

For the squeezed coherent states we have here, the conditional distribution is 
$\rho_{(\mathbf{p}|\mathbf{q})}=\mathcal{N}\left(\mathbf{p|\langle \mathbf{\hat{p}}\rangle}_{(\mathbf{p}|\mathbf{q})},\frac{\hbar}{2}\mathbf{s}^{-2}\right)$. Then, the variance of conditional momenta is $\frac{\hbar}{2}\mathbf{s}^{-2}$ and
\begin{eqnarray}\label{eq:Temperature_scalar}
k_B\mathbb{T}=\frac{\hbar}{2}{\rm{Tr}}\left(\mathbf{c}\mathbf{s}^{-2}\right).
\end{eqnarray}
Thus, we can write the conditional distribution in a Maxwellian manner as in the realm of the Maxwell--Boltzmann statistics, i.e.,

\begin{eqnarray}
\rho_{(\mathbf{p}|\mathbf{q})}=\frac{1}{\left(2\pi k_B\right)^{n/2}}\frac{1}{\sqrt{{\rm{det}}\left(\mathbf{c}^{-1}\mathbfcal{T}\right)}}\exp{\left[-\frac{1}{2k_B}\left(\mathbf{p}-{\langle\mathbf{\hat{p}}\rangle}_{(\mathbf{p}|\mathbf{q})}\right)^\intercal\left(\mathbfcal{T}^{-1}\mathbf{c}\right)\left(\mathbf{p}-{\langle\mathbf{\hat{p}}\rangle}_{(\mathbf{p}|\mathbf{q})}\right)\right]}.
\end{eqnarray}

\subsubsection{Internal energy, quantum potential and a conditional virial relation}\label{sec:Internal energy, quantum potential and a conditional virial relation}
Until now, we emphasized the importance of the measurements that are done with respect to the post--selection of positions. Let us now define an energy that is obtained by coarse--graining the Hamiltonian operator of the system over the momentum variables. For this, we will make use of the conditional averages introduced in Section~(\ref{sec:Marginal and conditional distributions}). We write the result as a functional of the classical Hamiltonian, $H\left(\mathbf{q}, \mathbf{p},t\right)$, i.e.,
\begin{eqnarray}
\mathbb{U}_{(\mathbf{p}|\mathbf{q})}\left(\mathbf{q},t\right)&:={\langle \hat{H}\rangle}_{(\mathbf{p}|\mathbf{q})}
=H\left(\mathbf{q},{\langle \mathbf{\hat{p}}\rangle}_{(\mathbf{p}|\mathbf{q})},t\right)+\underbrace{\frac{1}{2}k_B\mathbb{T}}_{\mathbb{U}_{\rm{kin.}}}.
\label{eq:conditional_energy}
\end{eqnarray}
Then, the conditional internal energy, $\mathbb{U}_{(\mathbf{p}|\mathbf{q})}$, is composed of  (i) a portion including the classical Hamiltonian functional that inputs the effective streamline  momenta of the flow of the system as its momenta variable, (ii) a pure quantum contribution with an energy term, $\mathbb{U}_{\rm{kin.}}=k_B\mathbb{T}/2$ analogous to the internal energy in classical kinetic theory for a single degree of freedom. In Sonego's work \cite{Sonego:1991} the term that corresponds to our $\mathbb{U}_{(\mathbf{p}|\mathbf{q})}$ is referred to as a ``local energy'' due to its dependence on the local position coordinates. Though, we should keep in mind that $\mathbb{U}_{(\mathbf{p}|\mathbf{q})}$ involves averages over the momentum variables which are already post--selected over positions.

Let us recall that in the de Broglie--Madelung--Bohm approach, it is the quantum potential that is responsible for the observed quantum behaviour of a system. In eq.~(\ref{eq:quant_pot_matrix}) of Section~(\ref{sec:Polar decomposition}), we derived it as $Q=\left(-\hbar ^2\boldsymbol{\nabla _q}^\intercal\mathbf{c}\boldsymbol{\nabla _q} R\right)/2R$ for the higher dimensional case. In order to find its value for a generic squeezed coherent state, we substitute the real amplitude, $R(\mathbf{q},t)$, given in eq.~(\ref{eq:ampl_sq_coh}) into its definition. The result follows as
\begin{eqnarray}\label{eq:quantum_potential_coherent}
Q=\frac{\hbar}{2}{\rm{Tr}}\left(\mathbf{c}\,\mathbf{s}^{-2}\right)-\frac{1}{2}\left(\mathbf{q}-\langle\mathbf{\hat{q}}\rangle \right)^\intercal \mathbf{s}^{-2}\mathbf{c}\,\mathbf{s}^{-2}\left(\mathbf{q}-\langle\mathbf{\hat{q}}\rangle\right).
\end{eqnarray}
There is a common perception in the literature that $Q\rightarrow 0$ should hold as $\hbar \rightarrow 0$. This impression follows from the fact that quantum potential is the only term that distinguishes the classical Hamilton--Jacobi equation from its quantum version. This is anticipated to be true both for the original derivation, eq.~(\ref{eq:Sch_real}), and for our derivation for a generic quadratic system in eq.~(\ref{eq:Sch_real_generic}). Therefore, $\hbar \rightarrow 0$ is expected to give the classical limit. However, such an expectation does not hold for a standard coherent state of a simple harmonic oscillator problem even in one dimensional, stationary case \cite{Barut:1990}. This discussion is usually overlooked in the literature except in a few studies. For example, in \cite{Holland:1995}, this confusion is argued in detail. It is shown that there exist different criteria to define a classical limit, though they often seem to contradict with each other. It is then resulted that certain states do not have classical limits as $\hbar \rightarrow 0$ and the authors suggest a method to properly define a classical limit \cite{Holland:1995}. We propose a different explanation here.

It is known that there are many interpretations of the quantum potential for different scenarios in the literature. Ours will be mostly aligned with the ones in \cite{Grossing:2009} and in \cite{Dennis:2015} with important differences. For example, the methodology and the set up of \cite{Grossing:2009} is very different than ours. However, the provided link between the quantum potential and the kinetic internal energy of the system is quite similar. Let us now discuss \cite{Dennis:2015}, in which the authors interpret $Q$ as the internal energy of a system for stationary states. For example, for a 3--dimensional simple harmonic oscillator, with frequency $\omega$ being the same for each degree of freedom, they obtain
\begin{eqnarray}\label{eq:quantum_potential_oscillator}
Q_{\omega}=\frac{3\hbar \omega}{2}-\frac{1}{2}m\omega ^2 |\mathbf{r}|^2,
\end{eqnarray}
where $\mathbf{r}$ is the position vector and $m$ is the mass. Their interpretation is that the internal energy is given by the ground state energy, $3\hbar \omega /2$ minus the potential energy. However, the problem of $\hbar$ not appearing in the so--called potential energy term brings us to the discussion of the previous paragraph. Can the two terms that appear on the right hand side of eq.~(\ref{eq:quantum_potential_oscillator}) be treated equally? Besides, the internal energy is known to be a coarse--grained object in thermodynamics. Thus, the value of $Q$ itself can not be expected to give the internal energy. According to us, it is rather the expectation value of $Q$ that should be interpreted as the internal energy.

As the quantum potential depends on the positions and on time only, ${\langle Q \rangle}_{(\mathbf{p}|\mathbf{q})}=Q$ and
\begin{eqnarray}
{\langle Q \rangle}={\langle Q \rangle}_{(\mathbf{q})}=\frac{\hbar}{4}{\rm{Tr}}\left(\mathbf{c}\,\mathbf{s}^{-2}\right).
\end{eqnarray}
Thus, following the value of temperature given in eq.~(\ref{eq:Temperature_scalar}), we obtain
\begin{eqnarray}
{\langle Q \rangle}=\frac{k_B\mathbb{T}}{2}=\mathbb{U}_{\rm{kin.}}.
\end{eqnarray}
This gives exactly the quantum mechanical internal energy in the form that it appears in the kinetic theory. Moreover, physically meaningful, measurable quantities are given by the average values of the operators. Thus one should expect ${\langle Q \rangle}\rightarrow 0$ as $\hbar \rightarrow 0$, which is the case here.

We also observe that the maximum value of the quantum potential is obtained at $\mathbf{q}={\langle \mathbf{\hat{q}} \rangle}$. This is where the Gaussian position distribution $\rho _{(\mathbf{q})}$ also peaks. Moreover, $Q_{\rm{max.}}=2{\langle Q \rangle}$. This was found ``interesting'' without further explanation in \cite{Nicacio:2020} in which the quantum potential and its mean are obtained for the Gaussian states. In fact, $Q$ in eq.~(\ref{eq:quantum_potential_coherent}) signals an object which is expanded around its maximum value. In classical thermodynamics, such extensions are usually introduced to study fluctuations of thermodynamic variables at equilibrium \cite{Landau:1980}. 

Let us now calculate the variance of the quantum potential, $\left\langle \left(\Delta Q\right)^2 \right\rangle=\left\langle \left(Q-\langle Q \rangle \right)^2 \right\rangle$ around its maximum value, i.e.,
\begin{eqnarray}
\left\langle \left(\Delta Q\right)^2 \right\rangle&=&\left\langle \left(\frac{dQ}{d\mathbf{q}}\Big|_{\mathbf{q}=\langle \mathbf{\hat{q}} \rangle}\left(\mathbf{q}-\langle \mathbf{\hat{q}} \rangle\right)+ \frac{d^2Q}{d\mathbf{q}^2}\Big|_{\mathbf{q}=\langle \mathbf{\hat{q}} \rangle}\left(\mathbf{q}-\langle \mathbf{\hat{q}} \rangle\right)^2\right)^2 \right\rangle \nonumber\\
&=&\left(\mathbf{s^{-2}cs^{-2}}\right)_{ij}\left(\mathbf{s^{-2}cs^{-2}}\right)_{kn}\left\langle \left(\mathbf{q}-\langle \mathbf{\hat{q}} \rangle\right)_i\left(\mathbf{q}-\langle \mathbf{\hat{q}} \rangle\right)_j\left(\mathbf{q}-\langle \mathbf{\hat{q}} \rangle\right)_k\left(\mathbf{q}-\langle \mathbf{\hat{q}} \rangle\right)_n\right\rangle \nonumber\\
&=&\frac{\hbar ^2}{4}\left(\mathbf{s^{-2}cs^{-2}}\right)_{ij}\left(\mathbf{s^{-2}cs^{-2}}\right)_{kn}\left(\mathbf{s^2}_{ij}\mathbf{s^2}_{kn}+\mathbf{s^2}_{ik}\mathbf{s^2}_{jn}+\mathbf{s^2}_{in}\mathbf{s^2}_{jk}\right) \nonumber\\
&=&k_B^2\left(\left[{\rm{Tr}}\mathbfcal{T}\right]^2+2{\rm{Tr}}\mathbfcal{T}^2\right)=3k_B^2\mathbb{T}^2-4k_B^2\sum\limits_{i<j} \lambda _i\lambda_j.
\end{eqnarray}
Here, the second line follows from the fact that the first order fluctuation term vanishes at $\mathbf{q}=\langle \mathbf{\hat{q}} \rangle$. The third line follows from the fact that the higher order moments of a Gaussian distribution can be written as a function of the variance due to the Isserlis Theorem \cite{Isserlis:1918}. Here, the variance in question is $\boldsymbol{\sigma_{qq}}=\frac{\hbar}{2}\mathbf{s}^2$. The last line follows from the temperature matrix given in eq.~(\ref{eq:Temperature}) with $\lambda_i$ being its eigenvalues. The variance of quantum potential is thus proportional to the square of the temperature similar to the mean--square fluctuation of the energy in classical thermodynamics. Thus, we suggest that $Q$ involves both the kinetic internal energy at equilibrium and the fluctuations around it. 

On the other hand, it is also known that for a system in thermodynamic equilibrium, a virial relation is satisfied if the system concurrently satisfies a hydrodynamic equilibrium. We based our thermodynamic equilibrium on the invariance of our phase space entropy. In addition, our system satisfies a hydrodynamic equilibrium due to the probability distributions satisfying the Fokker--Plank equations as given in Section~(\ref{sec:Fokker-Planck equation, probability fluxes and the continuity equation}). Note that the standard virial relation in quantum mechanics has two typical derivations: (i) the commutator proof which makes use of the invariance of the infinitesimal generator $i\left(\mathbf{\hat{q}}\mathbf{\hat{p}}+\mathbf{\hat{p}}\mathbf{\hat{q}}\right)/2$ and is found in many textbooks (cf. \cite{Schiff:1968}) that goes back to Finkelstein \cite{Finkelstein:1928}; (ii) the proof which makes use of the group of dilatations that was first given by Fock \cite{Fock:1930}. In the literature, the problems regarding both of these methods have been investigated in many studies and alternative derivations have been proposed \cite{Albeverio:1972,Weislinger:1974,Kalf:1976,Gersch:1979,Leinfelder:1981}. Our aim here, on the other hand, is to suggest a coarse--grained version of the virial relation applicable for the statistical hydrodynamic interpretation we presented here. 

Consider the quadratic system we have that is in thermodynamic equilibrium with its surroundings. Accordingly, we define a quantum virial relation as
\begin{eqnarray}\label{eq:virial}
2\mathbb{U}_{\rm{kin.}}=-\mathbb{U}_{\rm{pot.}}:=-\left \langle \left(\mathbf{q}-\langle \mathbf{\hat{q}}\rangle \right)^\intercal \boldsymbol{\mathcal{F}} \right \rangle,
\end{eqnarray}
where  $\boldsymbol{\mathcal{F}}=\nabla _{\mathbf{q}}Q$ is the quantum force.  Equation (\ref{eq:virial}) is different from its standard analogue in the sense that it represents an effective system. For example, the temperature and thus the $\mathbb{U}_{\rm{kin.}}$ term exist solely due to the conditional variance of the momenta. We must also add that the virial relation above incorporates solely the quantum effects by considering a potential energy term derived from the quantum potential only. One could in principle consider a more generic form of this virial relation which incorporates a classical kinetic energy and a classical potential in addition to their pure quantum analogues. We believe that for such an investigation one should consider a mixed quantum--classical phase space formalism, for example, as in \cite{Burghardt:2005}. 

While closing this section we should emphasise that up until now we investigated the energy coarse--grained over momenta. In order to obtain the internal energy coarse--grained over the entire phase space, one needs to consider the expectation value of the Hamiltonian operator obtained through the joint phase space distribution. Then, we obtain a phase space internal energy as
\begin{eqnarray}
\mathbb{U}_{(\mathbf{q},\mathbf{p})}:={\langle \hat{H}\rangle}
=
H\left({\langle \mathbf{\hat{q}}\rangle},{\langle \mathbf{\hat{p}}\rangle},t\right)+\mathbb{U}_{\rm{kin.}}+\frac{\hbar}{4}{\rm{Tr}}\left(\mathbf{a}\mathbf{s}^2\right)
-\frac{\hbar}{4}{\rm{Tr}}\left(2\mathbf{bgs^2}-\mathbf{cgs^2g}\right).\label{eq:joint_energy}
\end{eqnarray}
Here, again, the first term on the r.h.s is the classical contribution to the energy regarding the Hamiltonian functional that inputs the phase space
average of the positions and momenta. The second term is the quantum kinetic internal energy on account of the variance of the conditional momenta. Recall that if the shearing/lensing matrix, $\mathbf{g}$, is zero, then ${\langle\mathbf{\hat{p}}\rangle}_{(\mathbf{p}|\mathbf{q})}={\langle\mathbf{\hat{p}}\rangle}$. In that case, the only term that differentiates the global energy, $\mathbb{U}_{(\mathbf{q},\mathbf{p})}$, from a conditionally averaged local one, $\mathbb{U}_{(\mathbf{p}|\mathbf{q})}$, is the third term, $\hbar{\rm{Tr}}\left(\mathbf{a}\mathbf{s}^2\right)/4$, that is the energy contribution coming from the variance of positions. This term has no analogue in classical theory as one assumes no variance in classical positions in general, at least theoretically.

\subsubsection{Relationship between the kinetic internal energy and the Maslov index}\label{sec:Relationship between the internal energy and the Maslov index}
A more profound observation is the relationship of the conditional kinetic internal energy, $\mathbb{U}_{\rm{kin.}}$, to the Maslov index. The latter, denoted by $\mu$, is an (half)--integer valued map that is usually associated with the closed loops of the Lagrangian subspaces of a symplectic vector space \cite{Maslov:1972}. It is also interpreted as a topological invariant which gives the winding number for periodic systems \cite{Arnold:1967}. For example, for a harmonic oscillator in 1-dimension, 
\begin{eqnarray}
\oint p dq=E_n \mathscr{T}=2\pi \hbar\left(n+\frac{\mu}{4}\right),
\end{eqnarray}
where $E_n$ is the energy of the $n^{\rm{th}}$ energy level and $\mathscr{T}$ is the period of the oscillations, $\mu$ takes the value of 2.

Here, we will refer to an extended definition of the Maslov's formula for generic symplectic paths defined by linear symplectomorphisms \cite{Cappell:1994,deGosson:1997,Mcduff:1998}. In that case the symplectic phase space transformation matrix of the system is also periodic.  
Suppose that there exists a $2n\times 2n$ matrix $\mathbf{f(u)}$ which is both orthogonal and symplectic. Then $\mathbf{f(u)}$ has an $n$--dimensional unitary representation, $\mathbf{u}=\mathbf{x}+i\mathbf{y}$, with real $\mathbf{x}$ and $\mathbf{y}$, i.e., $\mathbf{u}(n)\sim Sp(2n)\cap O(2n)$. One then defines a Lagrangian subspace, $\Lambda \in \mathcal{L}(n)$, via
\begin{eqnarray}
\Lambda = \rm{Im}\left(
\begin{array}{c}
    \mathbf{x}  \\
    \mathbf{y} 
\end{array}
\right).
\end{eqnarray}
Now consider a loop $\Lambda (t)=\Lambda (t+\mathscr{T}) \in \mathcal{L}(n)$, where $\mathscr{T}=1$ is the normalized period of the system. Then a Maslov index can be defined for this loop as \cite{Mcduff:1998}
\begin{eqnarray}\label{eq:Maslov_symplectic}
\mu = \alpha (1)-\alpha (0),
\end{eqnarray}
where
\begin{eqnarray}
e^{i\pi \alpha (t)}=\rm{det}_{\mathbb{C}}\left[\mathbf{u}(t)\right], \qquad \mathbf{u}(t)= \left(\mathbf{x}(t)+i\mathbf{y}(t)\right), \qquad
\Lambda (t)=\rm{Im}\left(
\begin{array}{c}
    \mathbf{x}(t)  \\
    \mathbf{y}(t) 
\end{array}
\right), \,\,\, \forall \,t \in \mathbb{R}.
\end{eqnarray}
Thus, for a periodic linear system whose phase space transformations are governed by an orthogonal symplectic matrix, $\mathbf{f(u)}$, it is the change in the argument of the determinant of the corresponding unitary matrix, $\mathbf{u}$, that defines the Maslov index. 

In this paper, we consider systems that are not necessarily periodic in general. Therefore, the evolution of the expectation values of the position and momentum operators are governed by those symplectic matrices which are not necessarily orthogonal and periodic. In general, the energy is not a constant of time that can directly be related to the Maslov index. However, recall that in Section~(\ref{sec:Symplectic phase space of quadratic Hamiltonians}), we applied an Iwasawa factorization, eq.~(\ref{eq:Iwasawa}), to the governing symplectic matrix, $\mathbf{S}$. Its fractional Fourier transformer component is an orthogonal symplectic matrix and denoted by $\mathbf{f(u)}$. This matrix is the portion responsible for the generic rotations in phase space which are not necessarily around closed loops. Note that the matrix $\mathbf{f(u)}$ has a unitary representation, $\mathbf{u}$. Then, $\alpha (t)=\rm{arg}\left(\rm{det}\mathbf{u}\right)$ manifested itself in the phase function, $\mathcal{S}$, in eq.(\ref{eq:phase_sq_coh}), when we derived the generic squeezed coherent state wave function in its polar form. Later on, in Section~(\ref{sec:Consistency check: a pathway to thermodynamics}), we provided a consistency check which seemed redundant earlier on. Namely, we showed that in order for the squeezed coherent state wave function obtained from a Wigner distribution to uniquely match a wave function  which is obtained through the action of the metaplectic operators on a ground state (and which preserves its form as in eq.~(\ref{eq:Wavefn_Combescure}) throughout the evolution), the condition $d\alpha (t) /dt=\rm{Tr}\left(\mathbf{cs}^{-2}\right)$ has to be satisfied. Subsequently, the rate of change of the argument of the unitary matrix $\mathbf{u}$ is $\rm{Tr}\left(\mathbf{cs}^{-2}\right)$ which is defined in a similar fashion as the original Maslov index given in eq.~(\ref{eq:Maslov_symplectic}). Then, even though the system is not necessarily periodic, the change in $\alpha$ is a measure of the kinetic internal energy of the system, $\mathbb{U}_{\rm{kin.}}=\hbar\rm{Tr}\left(\mathbf{cs}^{-2}\right)/4$. This means that the generic phase space rotations, i.e., fractional Fourier transformations, are directly related to the quantum energy content of the semiclassical systems in any case. 

\section{Summary and conclusion}\label{sec:Summary and conclusion}
\begin{figure}
[htbp]
    \centering
    \includegraphics[scale=0.044]{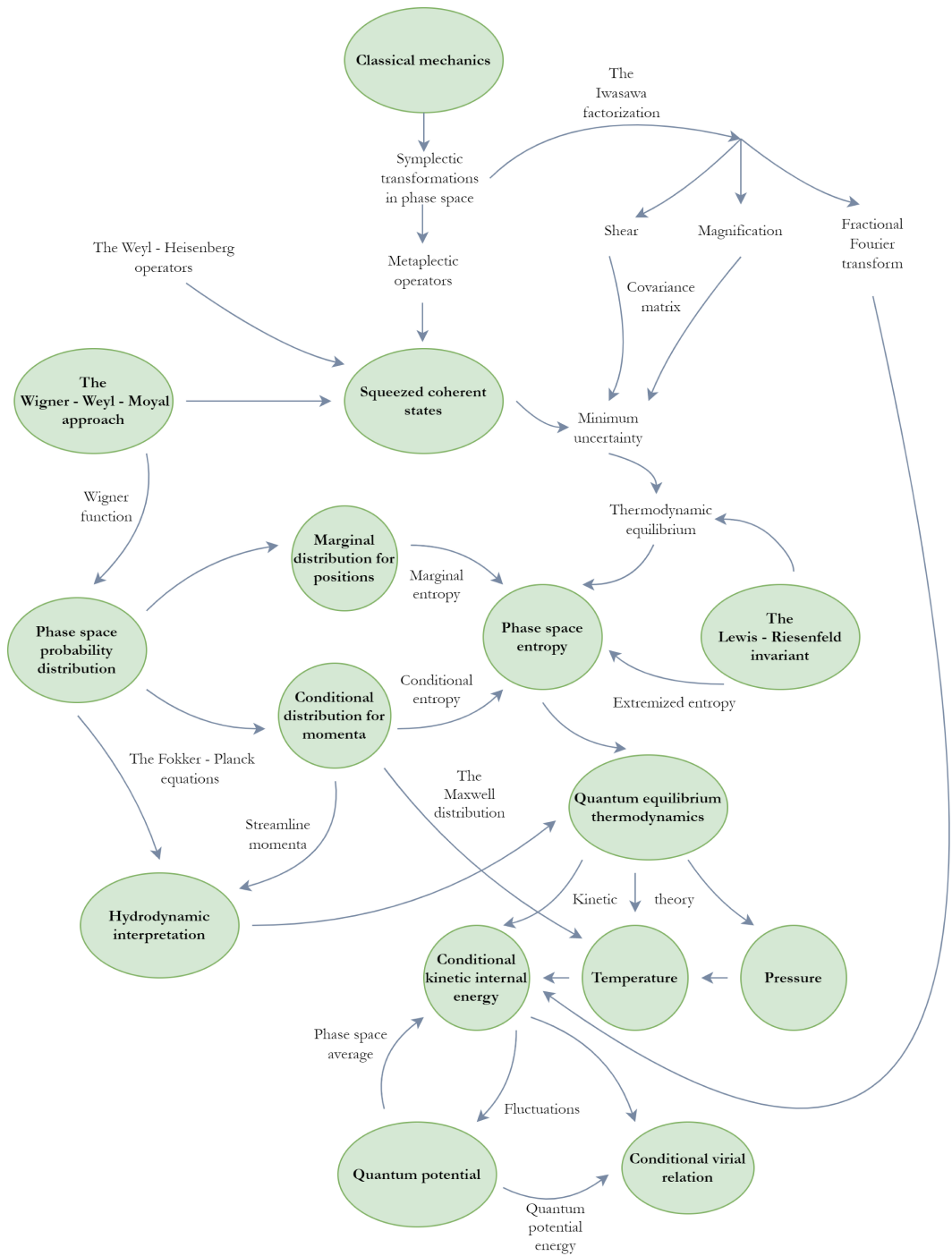}
    \caption{\label{fig:Mindmap}The conceptual relations relevant for the hydrodynamic interpretation of the generic squeezed coherent states outlined in this work.}
\end{figure}
Squeezed coherent states are mostly relevant for semi--classical physics or for systems whose Hamiltonian is in quadratic order with respect to the phase space coordinates. In our investigation, we studied squeezed coherent states in $n$--dimensions which are more generic than the standard ones. Those states can be exactly defined for systems with time dependent Hamiltonians. Moreover, the minimum uncertainty principle still holds at all times, without the uncertainty in positions and in momenta being necessarily equal.  

Within different interpretations of quantum mechanics, we chose to analyze the squeezed coherent states within a hydrodynamic interpretation which allowed us to investigate the system thermodynamically.  While doing that we placed the Wigner--Moyal--Weyl correspondence at the core of our construction as in \cite{Littlejohn:1985} where Littlejohn investigates the squeezed coherent states in detail. Eventually, we outlined a wide perspective by providing various links between the classical and the quantum mechanical paths in addition to highlighting certain statistical concepts that are mostly relevant for our hydrodynamic and thermodynamic analysis.
We summarize those conceptual links schematically in FIG~(\ref{fig:Mindmap}).

It is known that dynamics of a linear classical system is represented by linear symplectic transformations in a classical phase space. Such a system is driven by a quadratic Hamiltonian whose quantum analogue is also a quadratic function of position and momentum operators. In \cite{Littlejohn:1985}, Littlejohn derives the exact Gaussian form of the wave function of the squeezed coherent states by making use of: (i) the Weyl--Heisenberg operators which translate a state in the phase space (ii) the correspondance between the symplectic group and the metaplectic operators, the latter of which provide the spreading of the wave function. 

In order to investigate the squeezed coherent states within the hydrodynamic interpretation, we started by decomposing the corresponding wave function into its polar form as in the de Broglie--Madelung--Bohm approach. In the mean time, a phase space probability distribution was obtained via the Wigner function as in \cite{Littlejohn:1985}. Note that as our wave function in question is Gaussian, the associated Wigner function is non--negative and also Gaussian. This allows it to be a proper candidate for a phase space distribution function. 

Next, we started the thermodynamic analysis by defining a Shannon entropy via the Wigner function. For the case of the squeezed coherent states, this phase space Wigner entropy is a constant of time. Indeed, it taking a minimum value is consistent with the minimum of the Schr{\"{o}}dinger--Robinson uncertainty being satisfied by the squeezed coherent states. This is why we claimed that the system in question is in a dynamic thermodynamic equilibrium.

Further thermodynamic analysis was slightly more involved as it requires one to incorporate the statistical concepts with the quantum mechanical measurement process. For example, the hydrodynamic interpretation we followed was derived within the position representation. Momenta of the particles, on the other hand, are identified once the positions are selected. Therefore, if we would like to decompose a phase space distribution function into its portions involving the information about the positions and the momenta, we need to acknowledge the fact that the momenta are post--selected. Indeed, this fact was realized even in the early times of the statistical interpretation of quantum mechanics by Moyal \cite{Moyal:1949} and by Takabayasi \cite{Takabayasi:1954}.

Therefore, we treated the phase space distribution function associated with the squeezed coherent states as a joint distribution and decomposed it into two portions: (i) a marginal distribution for positions which is equal to the squared real amplitude of the wave function; (ii) a conditional distribution for momenta which are conditioned on positions. Note that our distribution functions are in Gaussian form and they all satisfy the Fokker--Planck equation exactly which we also discussed in detail. Moreover, we showed that the conditionally averaged momentum is equal to the so--called Bohm's momentum whose physical interpretation is still under debate. According to the hydrodynamic interpretation, Bohm's momentum reflects the streamline momentum of the particles that constitute the system. It being equal to the conditionally averaged momentum also shows that it is an effective object. This was also realized in certain other studies \cite{Sonego:1991} and its relation to the weak measurements of Aharonov \textit{et al.} \cite{Aharonov:1988} has been established before \cite{Hiley:2012w,Feyereisen:2015,Flack:2018}. 

Next, we returned back to the thermodynamic analysis via decomposing the phase space entropy into two portions by making use of: (i) the missing information contained in the positions which is obtained through the Shannon entropy of the marginal distribution; (ii) the missing information contained in the post--selected momenta which is obtained through the Shannon entropy of the conditional distribution of momenta. This is indeed analogous to the decomposition of the classical part of the Sackur--Tetrode entropy of the kinetic theory as discussed in the relevant section. 

After defining the probability distributions and their associated entropies, we followed Sonego's work \cite{Sonego:1991} in order to define a quantum pressure and a quantum temperature for the squeezed coherent states. Those satisfy an equation of state as in the classical kinetic theory. In the classical case, temperature is defined via the variance of the momentum which is sourced by the peculiar velocities of the particles with respect to the ensemble. Here, we explicitly showed that the quantum temperature is defined via the variance of the conditional momentum distribution in a similar fashion. This allowed us to rewrite our conditional distribution in a Maxwellian manner as in the Maxwell--Boltzmann distribution of the classical kinetic theory.

We further associated a conditional internal energy to our system in equilibrium. This conditional energy includes a contribution coming from the classical Hamiltonian which is modified by the conditionally averaged momenta and a portion that includes the quantum temperature term as in the form of the internal energy of the classical kinetic theory. Later on, we demonstrated the relationship of this kinetic internal energy term with the quantum potential of the de Broglie--Madelung--Bohm theory. According to us, the quantum potential includes the internal kinetic energy and its fluctuations around it at equilibrium. Accordingly, it is the expectation value of the quantum potential that gives the internal kinetic energy of the system. Eventually, we suggested a conditional virial relation that associates the kinetic internal energy of the system with a potential energy term sourced solely by the quantum potential.

In brief, our outcome is a quantum kinetic theory associated with $n$--dimensional squeezed coherent states in which the underlying thermodynamics is time dependent. This is unlike other works in the literature where the main idea behind the construction of quantum thermodynamics is usually adopted from the standard classical thermodynamics. Namely, the system is assumed to relax to an equilibrium in time and the energy of the system is kept constant throughout the evolution. Seemingly, the lack of time dependent quantum thermodynamic investigations follows from the lack of a time dependent equilibrium classical thermodynamics formalism constructed on a symplectic phase space. 

Certain delicate issues also caught our attention throughout our investigation. The first one follows from the Iwasawa factorization of the symplectic transformation matrix of the underlying classical system. This factorization essentially allows one to identify the lensing/shearing, pure magnification and the rotation--like portions of the phase space transformations. For the squeezed coherent states, those sub--transformations play essential roles in the quantum picture. For example, the square of the magnifications in classical positions manifests itself as a  quantum variance of the position uncertainty. Likewise, the square of the demagnification in classical momenta manifests itself as the quantum variance of the conditional momentum uncertainty. The matrix that is responsible for a shearing effect in the classical phase space also appears in the phase space covariance matrix. In addition, it is this shearing term that differentiates the conditionally averaged momenta from the phase space averaged ones. Recall that the former are equal to the real part of the weak measurements of momenta. To be more specific, when the underlying classical trajectory has zero shearing effect, the conditionally averaged momenta, or Bohm's momenta, are equal to the phase space expectation values of the momentum operator. Then, the covariance matrix takes a block diagonal form and one satisfies an exact Heisenberg uncertainty.

On the other hand, unlike the shears and the magnifications, it is the classical rotation--like transformation that manifests itself in the quantum kinetic internal energy of the system. To be more specific, there exists a unitary matrix representation of the fractional Fourier transformer part of the symplectic matrix that guides the classical evolution. It is the time rate of change of the argument of the determinant of this unitary matrix which tell us about the quantum kinetic energy content of a system. This is similar to the definition of the Maslov index characterized for the symplectic paths, which also contributes to the energy content of a system with periodic orbits.

As we indicated before, we considered only the squeezed coherent states of a linear system here. If we were to include the thermal states to our investigation, then we could still define a Gaussian Wigner function and a phase space probability distribution. However, in that case, the Wigner matrix in question would not be a symplectic matrix and its evolution in phase space would not be so trivial (cf. \cite{Yeh:1993}). This would result in a Shannon entropy of the phase space distribution which is not a constant of time, meaning the system would not be at its equilibrium. We believe this is a good point to start investigating the non--equilibrium thermodynamics of linear systems. In that case, the definition of the Wigner ellipsoid, the corresponding information entropy and the Mahalanobis distance of statistics \cite{Mahalanobis:1936} seem to be interconnected. We leave these issues for our future project.

Finally, in this work, we stayed within the linear regime only. Even though this can be seen somewhat restrictive, it is still relevant for certain application areas within the fields of quantum thermodynamics and quantum engines. Note that our results are directly relevant for experimental testing thanks to the developments in symplectic and optical tomography. The main idea behind tomographic methods is to reconstruct the classical and/or quantum phase space probability distributions via direct measurements. Even though those methods were initially formulated for somewhat restricted cases \cite{Bertrand:1987,Smithey:1993,Dariano:1995,Mancini:1997}, recent studies seem to be promising in terms of providing wider application areas \cite{Manko:2000,Healy:2015,Manko:2019,Manko:2021}. We suggest that tomography techniques, combined with the construction presented here, can in principle be used to analyse systems within a time dependent thermodynamic setting. Specifically, our results might find some area of use within the quantum optomechanical problems as in \cite{Zhang:2014,Qian:2012,Lorch:2014,Elouard:2015,Abari:2019,Bennett:2020} where stability and efficiency issues are open problems for time dependent systems. 

\section*{\label{Acknowledgements}Acknowledgements}
The author thanks Thomas Buchert, Jean-Pierre Gazeau, Przemys{\l{}}aw Ma{\l{}}kiewicz and Jan Jakub Ostrowski for their comments. 
\section*{\label{Funding}Funding}
This work is a part of a project that has received funding from the European Research Council (ERC) under the European Union's Horizon 2020 research and innovation programme (grant agreement ERC advanced Grant 740021ARTHUS, PI: Thomas Buchert).

%
%


\appendix

\section{Littlejohn's derivation of squeezed coherent states }\label{appendix:squeezed coherent states}
In \cite{Littlejohn:1985}, Littlejohn defines a
translation operator, $\hat{T}$, that is given within the Weyl--Heisenberg algebra. This operator is responsible for translating a given object. For instance, translation of $\hat{\mathbf{z}}$ by an amount $\mathbf{z'}$  is given by
\begin{eqnarray}
\hat{T}^\dag \left(\mathbf{z'}\,\right)\hat{\mathbf{z}}\,\hat{T} \left(\mathbf{z'}\right)=\hat{\mathbf{z}}+\mathbf{z'},
\end{eqnarray}
such that
\begin{eqnarray}
 \left\langle \Psi \right| \hat{T}^\dag \left(\mathbf{z'}\right)\hat{\mathbf{z}}\,\hat{T} \left(\mathbf{z'}\right)\left| \Psi \right\rangle = \mathbf{z}+\mathbf{z'}.
\end{eqnarray}
Formally, they are defined as 
\begin{eqnarray}
\hat{T}\left(\mathbf{z'}\right)&=&\exp\left(\frac{i}{\hbar}\hat{\omega}\left[\hat{\mathbf{z}},\mathbf{z'}\right]\right)\nonumber \\
&=&\exp\left(-\frac{i}{\hbar}\left(\mathbf{p'}\cdot\hat{\mathbf{q}}-\mathbf{q'}\cdot\hat{\mathbf{p}}\right)\right),
\end{eqnarray}
where the operator $\hat{\omega}\left[\hat{\mathbf{z}},\mathbf{z'}\right]$ acts as a quantum counterpart of the symplectic two--form ${\omega}\left({\mathbf{z}},\mathbf{z'}\right)$ given in eq.~(\ref{eq:symp_form}). The operator $\hat{T}$ is unitary and it satisfies
\begin{eqnarray}
\hat{T}^{-1}\left(\mathbf{z'}\right)=\hat{T}\left(-\mathbf{z'}\right)=\hat{T}^{\dag}\left(\mathbf{z'}\right).
\end{eqnarray}
However, its product rule follows as
\begin{eqnarray}\label{eq:T_product_rule}
\hat{T}\left(\mathbf{z}\right)\hat{T}\left(\mathbf{z'}\right)=\exp\left(\frac{i}{2\hbar}{\omega}\left({\mathbf{z}},\mathbf{{z}'}\right)\right)\hat{T}\left(\mathbf{z}+\mathbf{z'}\right),\nonumber \\
\end{eqnarray}
and thus those operators do not form a group due to the phase factor.
Nevertheless, they can be used to generate a propagator that evolves the quantum expectation values on the classical trajectory. 

Imagine segmenting the classical trajectory into $N$ straight pieces with a corresponding translation operator for each segment. By (i) making use of their products, i.e.,
\begin{eqnarray}
\hat{T}\left(\mathbf{z_N}-\mathbf{z_{N-1}}\right)...\,\hat{T}\left(\mathbf{z_2}-\mathbf{z_{1}}\right)\hat{T}\left(\mathbf{z_1}-\mathbf{z_{0}}\right), \nonumber \\
\end{eqnarray}
(ii) using the product rule in eq.~(\ref{eq:T_product_rule}), and (iii) considering the antisymmetry of the symplectic two--form, one can obtain a propagator corresponding to $\hat{T}$ as \cite{Littlejohn:1985},
\begin{eqnarray}\label{eq:U_T}
\hat{U}_{\hat{T}}(t)=\exp\left(-\frac{i}{2\hbar}\int_{0}^{t}\omega\left(\mathbf{z},\mathbf{\dot{z}}\right)dt\right)\hat{T}\left(\mathbf{z}(t)\right)\hat{T}^\dag\left(\mathbf{z_0}\right),\nonumber \\
\end{eqnarray}
in the limit $N \rightarrow \infty$ and $\mathbf{z_N}=\mathbf{z}(t)$. Here, the overdot denotes the standard total derivative with respect to time parameter. Note that the propagator $\hat{U}_{\hat{T}}(t)$ takes a Gaussian coherent state to another Gaussian coherent state with a different expectation value. It does not take care of the spreading of the wave function which is accommodated by the squeezing of the wave packet. For this, one introduces the metaplectic operators.

Within the set of all unitary automorphisms of $L^2\left(\mathbb{R}^n\right)$, the metaplectic group $Mp\left(2n,\mathbb{R}\right)$ is the subgroup generated by the quadratic Fourier transforms. One can show that there exists a continuous group epimorphism, a projection map, $\pi :Mp \rightarrow Sp$ that is the two--fold covering of the symplectic group \cite{deGosson:1997, deGosson:2006}. This means that for every symplectomorphism, i.e., a linear canonical transformation here, there exist two associated metaplectic unitary operators that differ by a sign. Then, only after specifying a symplectic matrix, $\mathbf{S}$, and a choice of a sign, $\sigma=\pm 1$, one can associate a metaplectic operator, $\hat{M}(\mathbf{S},\sigma)$, that forms a group. In general, a metaplectic operator satisfies
\begin{eqnarray}
\hat{M}(\mathbf{S_1})\hat{M}(\mathbf{S_2})=\pm \hat{M}(\mathbf{S_1}\mathbf{S_2}),\\
\hat{M}^{-1}(\mathbf{S})=\hat{M}^{\dag}(\mathbf{S})=\pm \hat{M}(\mathbf{S^{-1}}).
\end{eqnarray}
For our quadratic system, they propagate $\hat{\mathbf{z}}$ as the classical phase space vectors propagate, i.e.,
\begin{eqnarray}
\hat{M}^{\dag}(\mathbf{S})\hat{\mathbf{z}}\hat{M}(\mathbf{S})=\mathbf{S}\hat{\mathbf{z}}.
\end{eqnarray}
Given a metaplectic operator associated with the symplectic matrix $\mathbf{S}$, there exists a corresponding unitary operator $\hat{U}_{\hat{M}}$. In fact, this operator has been rediscovered by researchers in different fields many times and can be represented in the position space as

\begin{eqnarray}
\hat{U}_{\hat{M}}\left(\mathbf{q}, \mathbf{\tilde{q}}\right)
&=&\left\langle \hat{\mathbf{q}}(q) \right| \hat{M} \left| \hat{\mathbf{q}}(\tilde{q}) \right\rangle \nonumber\\
&=&\frac{\sigma}{\left(2i\pi\hbar\right)^{n/2}\sqrt{\left|\rm{det}\mathbf{B}\right|}}
\exp {\left[\frac{i}{2\hbar}\left(\mathbf{q}^\intercal \mathbf{DB}^{-1}\mathbf{q}-2\mathbf{\tilde{q}}^\intercal \mathbf{B}^{-1}\mathbf{q}+\mathbf{\tilde{q}}^\intercal \mathbf{B}^{-1}\mathbf{A}\mathbf{\tilde{q}}\right)\right]},
\end{eqnarray}

so that this unitary operator acts as a propagator for  the wave function in position representation, i.e.,
\begin{eqnarray}
\psi\left(\mathbf{q}\right)=\int d\mathbf{\tilde{q}}\,\hat{U}_{\hat{M}}\left(\mathbf{q}, \mathbf{\tilde{q}}\right)\psi'\left(\mathbf{\tilde{q}}\right),
\end{eqnarray}
where the integral is taken from $-\infty$ to $+\infty$ as in other integrals that appear without indicated limits in this work.
In order to understand the effect of metaplectic operators acting on translated states one uses the property 
\begin{eqnarray}
\hat{M}^{\dag}(\mathbf{S})\hat{T}\left(\mathbf{z'}\right)\hat{M}(\mathbf{S})=\hat{T}\left(\mathbf{S}^{-1}\mathbf{z'}\right),
\end{eqnarray}
which is paramount for the investigation of the squeezed coherent states.

From now on, let $ \left| \mathbf{0} \right\rangle$ represent the ground state of a coherent state with $\left\langle \mathbf{0} \left| \hat{\mathbf{z}} \right| \mathbf{0} \right\rangle =\mathbf{0}$. Also consider a coherent state $\left| \mathbf{z} \right\rangle$ with expectation value $\langle \hat{\mathbf{z}} \rangle$ by keeping in mind that $ \left| \mathbf{z} \right\rangle=\hat{T}\left(\mathbf{z}\right)\left| \mathbf{0}\right\rangle$. Then, the combined effect of the metaplectic and the Weyl--Heisenberg operators on a coherent ground state is given by
\begin{eqnarray}
\hat{M}\left(\mathbf{S}\right)\hat{T}\left( \mathbf{z} \right)\left| \mathbf{0}\right\rangle&=&\hat{T}\left(\mathbf{S} \mathbf{z} \right)\hat{M}(\mathbf{S})\left| \mathbf{0}\right\rangle \nonumber \\
&=&\hat{M}(\mathbf{S})\left| \mathbf{z}\right\rangle.
\end{eqnarray}
Next, it can be shown that
there exists a well defined unitary operator in $L^2\left(\mathbb{R}^n\right)$ that propagates an initial state $\left| \mathbf{z_0}\right\rangle$ by
\begin{eqnarray}\label{eq:U_prop}
\hat{U}\left(t,\mathbf{z_0}\right)=\exp{\left(\frac{i\gamma (t)}{\hbar}\right)}\hat{T}\left( \mathbf{z}(t)\right)\hat{M}\left(\mathbf{S}(t)\right)\hat{T}^\dag \left(\mathbf{z_0}\right).\nonumber \\
\end{eqnarray}
The propagator $\hat{U}\left(t,\mathbf{z_0}\right)$ accommodates the action of both the metaplectic and the translation operators such that 
\begin{eqnarray}\label{eq:psi_U_psi_0}
\left| \psi\right\rangle=\hat{U}\left(t,\mathbf{z_0}\right)\left| \psi_0\right\rangle,
\end{eqnarray}
for an arbitrary initial state $\left| \psi_0\right\rangle$.
Here, 
\begin{eqnarray}\label{eq:Phase_Littlejohn}
\gamma\left(t\right)=\frac{1}{2}\int_0^t dt \left[\mathbf{p}\mathbf{\dot{q}}-\mathbf{q}\mathbf{\dot{p}}-2H\left(\mathbf{z},t\right)\right]
\end{eqnarray}
is a phase factor that does not necessarily appear in all of the coherent state propagators in the literature. As we follow Littlejohn's argument in \cite{Littlejohn:1985} we choose to include this phase factor so that the definition of translation operators can be extended to have a group property.

It is now easy to show that the unitary operator, $\ref{eq:U_prop}$, satisfies the Schr{\"{o}}dinger equation \ref{eq:Sch_U_prop}. It can be used to derive the squeezed coherent state wave function, eq.~(\ref{eq:psi_seq_coh}), from the ground state wave function, eq.~(\ref{eq:psi_ground}), as explained in Section~(\ref{sec:Squeezed coherent states}).

\section{Classical and quantum invariants in relation to the Wigner function}\label{appendix:Invariants}
Consider a Hamiltonian operator for a harmonic oscillator with a unit mass and time dependent frequency, $\Omega (t)$, in 1--dimension as
\begin{eqnarray}\label{eq:Ham_harm_osc}
\hat{H}\left(\hat{q},\hat{p};t\right)=\frac{1}{2}\hat{p}^2+[\Omega  (t)]^2\hat{q}^2.
\end{eqnarray}
Lewis \cite{Lewis:1967} and Lewis$\&$Riesenfeld \cite{LewisRiesenfeld:1969} define a dynamic invariant operator, $\hat{I}=\hat{I}(t)$, such that 
\begin{eqnarray}\label{eq:invar_op}
\dot{\hat{I}}=\frac{d\hat{I}}{dt}=\frac{\partial \hat{I}}{\partial t}+\frac{1}{i\hbar}[\hat{I},\hat{H}].
\end{eqnarray}
This invariant operator is given by \cite{Lewis:1967, LewisRiesenfeld:1969}
\begin{eqnarray}
\hat{I}=\frac{1}{2}\left(\lambda \frac{\hat{q}^2}{\zeta^{2}}+\left[\zeta\hat{p}-\dot{\zeta}\hat{q}\right]^2\right)
\end{eqnarray}
provided that the complex variable $\zeta=\zeta(t)$ satisfies the Ermakov equation
\begin{eqnarray}\label{eq:Ermakov}
\Ddot{\zeta}+[\Omega  (t)]^2\zeta=\frac{\lambda}{\zeta^3},
\end{eqnarray}
where $\lambda$ is a constant.

In \cite{Yeh:1993}, Yeh shows that a generalized Lewis--Riesenfeld invariant operator
can be defined for a time dependent quadratic Hamiltonian following the Weyl correspondence of the Wigner ellipsoid. Let us consider a system associated with Gaussian states. Then, there exists a classical invariant
\begin{eqnarray}\label{eq:Wigner_invariant}
I=\frac{1}{2}\left(\mathbf{z}-\left\langle \hat{\mathbf{z}} \right \rangle\right)^\intercal. \mathbf{W}.\left(\mathbf{z}-\left\langle \hat{\mathbf{z}} \right \rangle\right),
\end{eqnarray}
associated with the system that has a corresponding quantum invariant operator 
\begin{eqnarray}\label{eq:Wigner_invariant_op}
\hat{I}=\frac{1}{2}\left(\hat{\mathbf{z}}-\left\langle \hat{\mathbf{z}} \right \rangle\right)^\intercal. \mathbf{W}.\left(\hat{\mathbf{z}}-\left\langle \hat{\mathbf{z}} \right \rangle\right),
\end{eqnarray}
which satisfies $d\hat{I}/dt=0$ as in eq.~(\ref{eq:invar_op}). This follows from the fact that not only the classical phase space coordinates and momenta but also their quantum correspondences evolve via linear symplectomorphisms for quadratic Hamiltonians. Then, the time evolution of $\hat{\mathbf{z}}(t)=\mathbf{S}\hat{\mathbf{z}}(0)$ and $\left\langle \hat{\mathbf{z}} \right \rangle(t)=\mathbf{S}\left\langle \hat{\mathbf{z}} \right \rangle(0)$ cancels the time evolution of the Wigner matrix, $\mathbf{W}(t)=\mathbf{S}^{-\intercal}\mathbf{W}(0)\mathbf{S}^{-1}$, in the operator $\hat{I}$ above. Indeed, one can conclude that for the squeezed coherent states of the quadratic Hamiltonians, invariance of the generalized Lewis--Riesenfeld operator follows from the correspondance between the invariant Wigner ellipsoid and the density operator \cite{Yeh:1993}. 

Note that there exist certain other generalizations of the Lewis--Riesenfeld invariant method. For example, in \cite{Malkin:1969,Malkin:1973,Dodonov:1975,Dodonov:2003}, the authors make use of the Green function method in order to identify invariant integrals of motion. Those are sometimes referred to as the Malkin--Man'ko invariants in the literature. In others, connections to Feynmann propagators \cite{Dhara:1984} and classical and quantum equations of motions have been found \cite{Bertin:2012}. We believe, those generalizations can help one to identify the invariants of more generic quantum systems in the future in relation to phase space techniques \cite{Abdalla:2005} and symplectic transformations.

 \bibliographystyle{elsarticle-num} 
 \bibliography{references}

\providecommand{\noopsort}[1]{}\providecommand{\singleletter}[1]{#1}
\begin{thebibliography}{100}
\expandafter\ifx\csname url\endcsname\relax
  \def\url#1{\texttt{#1}}\fi
\expandafter\ifx\csname urlprefix\endcsname\relax\def\urlprefix{URL }\fi
\expandafter\ifx\csname href\endcsname\relax
  \def\href#1#2{#2} \def\path#1{#1}\fi

\bibitem{Schrodinger:1926}
E.~Schr{\"{o}}dinger, {Der stetige Ubergang von der Mikro-zur Makromechanik},
  Naturwiss. 14 (1926) 664--666.

\bibitem{Glauber:1963tx}
R.~J. Glauber, {Coherent and incoherent states of the radiation field}, Phys.
  Rev. 131 (1963) 2766--2788.

\bibitem{Glauber:1963fi}
R.~J. Glauber, {The Quantum theory of optical coherence}, Phys. Rev. 130 (1963)
  2529--2539.

\bibitem{Perelomov:1986}
A.~M. Perelomov, A.~M. Perelomov, Generalized Coherent States and Their
  Applications, Springer-Verlag, Berlin Heidelberg, 1986.

\bibitem{Littlejohn:1985}
R.~G. Littlejohn, {The Semiclassical Evolution of Wave Packets}, Phys. Rept.
  138 (1986) 193.

\bibitem{Folland:1989}
G.~Folland, Harmonic Analysis in Phase Space, Princeton University Press, New
  Jersey, 1989.

\bibitem{Caves:1981}
C.~M. Caves, Quantum-mechanical noise in an interferometer, Phys. Rev. D 23
  (1981) 1693--1708.

\bibitem{Walls:1983}
D.~F. Walls, Squeezed states of light, Nature 306~(5939) (1983) 141--146.

\bibitem{Wu:1987}
L.-A. Wu, M.~Xiao, H.~J. Kimble, Squeezed states of light from an optical
  parametric oscillator, J. Opt. Soc. Am. B 4~(10) (1987) 1465--1475.

\bibitem{Ritze:1987}
H.-H. Ritze, A.~Bandilla, Squeezing and first-order coherence, J. Opt. Soc. Am.
  B 4~(10) (1987) 1641--1644.

\bibitem{Tse:2019}
M.~Tse~\textit{et al.}, Ligo-Collaboration, Quantum-enhanced advanced ligo
  detectors in the era of gravitational-wave astronomy, Phys. Rev. Lett. 123
  (2019) 231107.

\bibitem{Acernese:2019}
F.~Acernese~\textit{et al.}, Virgo-Collaboration, Increasing the astrophysical
  reach of the advanced virgo detector via the application of squeezed vacuum
  states of light, Phys. Rev. Lett. 123 (2019) 231108.

\bibitem{Lough:2021}
J.~Lough~\textit{et al.}, GEO600-Collaboration, First demonstration of 6 db
  quantum noise reduction in a kilometer scale gravitational wave observatory,
  Phys. Rev. Lett. 126 (2021) 041102.

\bibitem{Li:2020}
T.~Li, F.~Li, C.~Altuzarra, A.~Classen, G.~S. Agarwal, Squeezed light induced
  two-photon absorption fluorescence of fluorescein biomarkers,
  Appl.~Phys.~Lett. 116~(25) (2020) 254001.

\bibitem{Lawrie:2020}
B.~Lawrie, R.~Pooser, P.~Maksymovych, Squeezing noise in microscopy with
  quantum light, Trends~Chem. 2~(8) (2020) 683--686.

\bibitem{Grishchuk:1990bj}
L.~P. Grishchuk, Y.~V. Sidorov, {Squeezed quantum states of relic gravitons and
  primordial density fluctuations}, Phys. Rev. D 42 (1990) 3413--3421.

\bibitem{Albrecht:1992kf}
A.~Albrecht, P.~Ferreira, M.~Joyce, T.~Prokopec, {Inflation and squeezed
  quantum states}, Phys. Rev. D 50 (1994) 4807--4820.

\bibitem{Martin:2012pea}
J.~Martin, V.~Vennin, P.~Peter, {Cosmological Inflation and the Quantum
  Measurement Problem}, Phys. Rev. D 86 (2012) 103524.

\bibitem{Bohm:1952_I}
D.~Bohm, A suggested interpretation of the quantum theory in terms of ``hidden"
  variables. i, Phys. Rev. 85 (1952) 166--179.

\bibitem{Bohm:1952_II}
D.~Bohm, A suggested interpretation of the quantum theory in terms of ``hidden"
  variables. ii, Phys. Rev. 85 (1952) 180--193.

\bibitem{DeBroglie:1925}
L.~{de Broglie}, Recherches sur la th\'eorie des quanta, Ann. Phys. 10~(3)
  (1925) 22--128.

\bibitem{DeBroglie:1927}
L.~{de Broglie}, {La m{\'e}canique ondulatoire et la structure atomique de la
  mati{\`e}re et du rayonnement}, J.~Phys.~Rad. 8~(5) (1927) 225--241.

\bibitem{Madelung:1927}
E.~{Madelung}, {Quantentheorie in hydrodynamischer Form}, Z.~Phys. 40~(3--4)
  (1927) 322--326.

\bibitem{Takabayasi:1954}
T.~Takabayasi, {The Formulation of Quantum Mechanics in terms of Ensemble in
  Phase Space}, Prog.~Theor.~Phys. 11~(4-5) (1954) 341--373.

\bibitem{Wigner:1932}
E.~Wigner, On the quantum correction for thermodynamic equilibrium, Phys. Rev.
  40 (1932) 749--759.

\bibitem{Moyal:1949}
J.~E. Moyal, Quantum mechanics as a statistical theory,
  Math.~Proc.~Cambridge~Phil.~Soc. 45~(1) (1949) 99–124.

\bibitem{Hiley:2015xy}
B.~J. Hiley, {On the relationship between the Wigner--Moyal approach and the
  quantum operator algebra of Von Neumann}, J.~Comput.~Electron. 14~(4) (2015)
  869--878.

\bibitem{Lewis:1967}
H.~R. Lewis, Classical and quantum systems with time-dependent
  harmonic-oscillator-type hamiltonians, Phys. Rev. Lett. 18 (1967) 510--512.

\bibitem{LewisRiesenfeld:1969}
J.~{Lewis}, H.~R., W.~B. {Riesenfeld}, {An Exact Quantum Theory of the
  Time-Dependent Harmonic Oscillator and of a Charged Particle in a
  Time-Dependent Electromagnetic Field}, J.~Math.~Phys. 10~(8) (1969)
  1458--1473.

\bibitem{Hiley:2006}
B.~J. Hiley, Beyond the Quantum, World Scientific Publishing, Singapore, 2006,
  Ch. Phase Space Description of Quantum Mechanics and Non--Commutative
  Geometry: Wigner–Moyal and Bohm in a Wider Context, pp. 203--211.

\bibitem{Hiley:2010}
B.~J. Hiley, {On the Relationship Between the Wigner--Moyal and Bohm Approaches
  to Quantum Mechanics: A Step to a More General Theory?}, Found.~Phys. 40
  (2010) 356--367.

\bibitem{Hiley:2012w}
B.~J. Hiley, Weak values: Approach through the clifford and moyal algebras,
  J.~Phys.: Conf. Ser. 361 (2012) 012014.

\bibitem{Colomes:2015}
E.~Colom{\'e}s, Z.~Zhan, X.~Oriols, Comparing wigner, husimi and bohmian
  distributions: which one is a true probability distribution in phase space?,
  J.~Comput.~Electron. 14~(4) (2015) 894--906.

\bibitem{Barut:1990}
A.~Barut, M.~Bozic, {The quantum potential and causal trajectories for
  stationary states and for coherent states}, Ann.~Fond.~Louis de Broglie
  15~(1) (1990) 67--90.

\bibitem{Dey:2013}
S.~Dey, A.~Fring, Bohmian quantum trajectories from coherent states, Phys. Rev.
  A 88 (2013) 022116.

\bibitem{Durr:2010}
D.~D{\"u}rr, S.~R{\"o}mer, On the classical limit of bohmian mechanics for
  hagedorn wave packets, J.~Funct.~Anal. 259~(9) (2010) 2404 -- 2423.

\bibitem{Yeh:1993}
L.~Yeh, Ermakov--lewis invariant from the wigner function of a squeezed
  coherent state, Phys. Rev. A 47 (1993) 3587--3592.

\bibitem{Sonego:1991}
S.~Sonego, Interpretation of the hydrodynamical formalism of quantum mechanics,
  Found.~Phys. 21~(10) (1991) 1135--1181.

\bibitem{Grossing:2009}
G.~Gr{\"{o}}ssing, On the thermodynamic origin of the quantum potential,
  Phys.~A: Stat.~Mech.~Appl. 388~(6) (2009) 811 -- 823.

\bibitem{Dennis:2015}
G.~{Dennis}, M.~A. {de Gosson}, B.~J. {Hiley}, Bohm's quantum potential as an
  internal energy, Phys.~Lett. A 379~(18) (2015) 1224--1227.

\bibitem{Bohm:1984}
D.~Bohm, B.~J. Hiley, Measurement understood through the quantum potential
  approach, Found.~Phys. 14~(3) (1984) 255--274.

\bibitem{Bohm:1995}
D.~Bohm, B.~Hiley, The undivided universe: an ontological interpretation of
  quantum theory, Routledge, London, 1995.

\bibitem{Gonzalez:2012}
P.~Gonz\'alez-D\'\i{}az, A.~Rozas-Fern\'andez, {Applied Bohmian Mechanics}, CRC
  Press, Boca Raton, 2012, Ch. {Subquantum Accelerating Universe}, pp.
  507--560.

\bibitem{Iwasawa:1949}
K.~{Iwasawa}, {On Some Types of Topological Groups}, Ann.~Math. 50~(3) (1949)
  507.

\bibitem{Arvind:1995}
{Arvind}, B.~Dutta, N.~Mukunda, R.~Simon, The real symplectic groups in quantum
  mechanics and optics, Pramana 45~(6) (1995) 471--497.

\bibitem{Wolf:2004}
K.~B. {Wolf}, {Geometric Optics on Phase Space}, Springer, Berlin Heidelberg,
  2004.

\bibitem{Bastiaans:1979}
M.~J. Bastiaans, Wigner distribution function and its application to
  first-order optics, J. Opt. Soc. Am. 69~(12) (1979) 1710--1716.

\bibitem{deGosson:2019arxiv}
M.~D. Gosson, Symplectic coarse-grained dynamics: Chalkboard motion in
  classical and quantum mechanics, arXiv: 1901.06554 (2019).

\bibitem{Husimi:1940}
K.~Husimi, Some formal properties of the density matrix,
  Proc.~Phys.~Math.~Soc.~Jpn. 22~(4) (1940) 264--314.

\bibitem{Combescure:2012}
M.~Combescure, D.~Robert, Coherent States and Applications in Mathematical
  Physics, Springer, Dordrecht, 2012.

\bibitem{Glauber:1963}
R.~J. Glauber, Coherent and incoherent states of the radiation field, Phys.
  Rev. 131 (1963) 2766--2788.

\bibitem{Sudarshan:1963}
E.~C.~G. Sudarshan, Equivalence of semiclassical and quantum mechanical
  descriptions of statistical light beams, Phys.~Rev.~Lett. 10 (1963) 277--279.

\bibitem{Kiesel:2013}
T.~Kiesel, Classical and quantum-mechanical phase-space distributions,
  Phys.~Rev.~A 87 (2013) 062114.

\bibitem{Manko:2021}
O.~V. Man'ko, V.~I. Man'ko, Probability representation of quantum states,
  Entropy 23~(5) (2021).

\bibitem{Santos:2017}
J.~P. Santos, G.~T. Landi, M.~Paternostro, Wigner entropy production rate,
  Phys. Rev. Lett. 118 (2017) 220601.

\bibitem{Brunelli:2018}
M.~Brunelli, L.~Fusco, R.~Landig, W.~Wieczorek, J.~Hoelscher-Obermaier,
  G.~Landi, F.~L. Semi{\~ao}, A.~Ferraro, N.~Kiesel, T.~Donner, Experimental
  determination of irreversible entropy production in out-of-equilibrium
  mesoscopic quantum systems, Phys. Rev. Lett. 121 (2018) 160604.

\bibitem{Malouf:2019}
W.~T.~B. Malouf, J.~P. Santos, L.~A. Correa, M.~Paternostro, G.~T. Landi,
  Wigner entropy production and heat transport in linear quantum lattices,
  Phys. Rev. A 99 (2019) 052104.

\bibitem{Belenchia:2020}
A.~{Belenchia}, L.~{Mancino}, G.~T. {Landi}, M.~{Paternostro}, {Entropy
  production in continuously measured Gaussian quantum systems},
  Npj~Quantum~Inf. 6 (2020) 97.

\bibitem{Adesso:2012}
G.~Adesso, D.~Girolami, A.~Serafini, Measuring gaussian quantum information and
  correlations using the r{\'e}nyi entropy of order 2, Phys. Rev. Lett. 109
  (2012) 190502.

\bibitem{Birula:1975}
I.~{Bia{\l}ynicki-Birula}, J.~{Mycielski}, {Uncertainty relations for
  information entropy in wave mechanics}, Commun.~Math.~Phys. 44~(2) (1975)
  129--132.

\bibitem{Birula:2011}
I.~Bialynicki-Birula, {\L}.~Rudnicki, Entropic Uncertainty Relations in Quantum
  Physics, Springer, Dordrecht, 2011, Ch.~1, pp. 1--34.

\bibitem{deGosson:2009}
M.~A. {de Gosson}, {The Symplectic Camel and the Uncertainty Principle: The Tip
  of an Iceberg?}, Found.~Phys. 39~(2) (2009) 194--214.

\bibitem{deGosson:2019}
N.~C. Dias, M.~A. de~Gosson, J.~N. Prata, A refinement of the
  robertson-schrödinger uncertainty principle and a hirschman-shannon
  inequality for wigner distributions, J.~Fourier~Anal.~Appl. 25~(1) (2019)
  210—241.

\bibitem{Dodonov:1980nx}
V.~V. Dodonov, E.~V. Kurmyshev, V.~I. Manko, {Generalized uncertainty relation
  and correlated coherent states}, Phys.~Lett.~A 79 (1980) 150--152.

\bibitem{deGosson:2006}
M.~A. de~Gosson, Symplectic Geometry and Quantum Mechanics, Birkh{\"a}user,
  Basel, 2006.

\bibitem{deGossonLuef:2009}
M.~A. {de Gosson}, F.~Luef, Symplectic capacities and the geometry of
  uncertainty: The irruption of symplectic topology in classical and quantum
  mechanics, Phys.~Rep. 484~(5) (2009) 131--179.

\bibitem{deGosson:2013}
M.~A. {de Gosson}, {Quantum Blobs}, Found.~Phys. 43~(4) (2013) 440--457.

\bibitem{deGosson:2013b}
M.~A. de~Gosson, The symplectic egg in classical and quantum mechanics,
  Am.~J.~Phys. 81~(5) (2013) 328--337.

\bibitem{Gromov:1985}
M.~Gromov, Pseudo holomorphic curves in symplectic manifolds., Invent.~Math. 82
  (1985) 307--348.

\bibitem{Deotto:1997}
E.~Deotto, G.~C. Ghirardi, {Bohmian mechanics revisited}, Found. Phys. 28
  (1998) 1--30.

\bibitem{Wiseman:2007}
H.~M. {Wiseman}, {Grounding Bohmian mechanics in weak values and bayesianism},
  New J.~Phys. 9~(6) (2007) 165.

\bibitem{Aharonov:1988}
Y.~Aharonov, D.~Z. Albert, L.~Vaidman, How the result of a measurement of a
  component of the spin of a spin-1/2 particle can turn out to be 100,
  Phys.~Rev.~Lett. 60 (1988) 1351--1354.

\bibitem{Dressel:2010}
J.~Dressel, S.~Agarwal, A.~N. Jordan, Contextual values of observables in
  quantum measurements, Phys. Rev. Lett. 104 (2010) 240401.

\bibitem{Dressel:2012}
J.~Dressel, A.~N. Jordan, Contextual-value approach to the generalized
  measurement of observables, Phys. Rev. A 85 (2012) 022123.

\bibitem{Dressel:2012b}
J.~Dressel, A.~N. Jordan, Significance of the imaginary part of the weak value,
  Phys. Rev. A 85 (2012) 012107.

\bibitem{Feyereisen:2015}
M.~R. {Feyereisen}, {How the Weak Variance of Momentum Can Turn Out to be
  Negative}, Found.~Phys. 45~(5) (2015) 535--556.

\bibitem{Flack:2018}
R.~Flack, B.~J. Hiley, Feynman paths and weak values, Entropy 20~(5) (2018).

\bibitem{Einstein:1905}
A.~Einstein, {\"{U}}ber die von der molekularkinetischen theorie der
  w{\"{a}}rme geforderte bewegung von in ruhenden fl{\"{u}}ssigkeiten
  suspendierten teilchen, Ann.~Phys. 322~(8) (1905) 549--560.

\bibitem{Einstein:1956}
A.~Einstein, Investigations on the Theory of the Brownian Movement, Dover
  Publications, New York, 1956.

\bibitem{Einstein:1990}
D.~Cahan, The collected papers of albert einstein. vol. 2, the swiss years:
  Writings, 1900--1909. john stachel, editor. david c. cassidy, j{\"u}rgen
  renn, and robert schulmann, associate editors. princeton university press,
  princeton, nj, 1989., Science 248~(4957) (1990) 878--879.

\bibitem{Bohm:1989}
D.~{Bohm}, B.~J. {Hiley}, {Non-locality and locality in the stochastic
  interpretation of quantum mechanics}, Phys.~Rep. 172~(3) (1989) 93--122.

\bibitem{Einstein:2006}
T.~Damour, O.~Darrigol, V.~Rivasseau, Einstein, 1905-2005: Poincar{\'e} Seminar
  2005, Birkh{\"a}user, Basel, 2006.

\bibitem{Ryskin:1997}
G.~Ryskin, Simple procedure for correcting equations of evolution: Application
  to markov processes, Phys. Rev. E 56 (1997) 5123--5127.

\bibitem{Sackur:1911}
O.~Sackur, Die anwendung der kinetischen theorie der gase auf chemische
  probleme, Ann.~Phys. 341~(15) (1911) 958--980.

\bibitem{Tetrode:1912}
H.~Tetrode, {Berichtigung zu meiner Arbeit: Die chemische Konstante der Gase
  und das elementare Wirkungsquantum}, Ann.~Phys. 344~(11) (1912) 255--256.

\bibitem{Ben:2008}
A.~Ben-Naim, A Farewell to Entropy: Statistical Thermodynamics Based on
  Information, World Scientific Publishing Company, Singapore, 2008.

\bibitem{Holland:1995}
P.~Holland, The Quantum Theory of Motion: An Account of the de Broglie-Bohm
  Causal Interpretation of Quantum Mechanics, Cambridge University Press,
  Cambridge, 1995.

\bibitem{Nicacio:2020}
F.~Nicacio, F.~T. Falciano, {Mean Value of the Quantum Potential and
  Uncertainty Relations}, Phys. Rev. A 101~(5) (2020) 052105.

\bibitem{Landau:1980}
L.~Landau, E.~Lifshitz, Statistical Physics, Vol.~5, Pergamon Press, Oxford,
  1980.

\bibitem{Isserlis:1918}
L.~Isserlis, On a formula for the product-moment coefficient of any order of a
  normal frequency distribution in any number of variables, Biometrika 12~(1/2)
  (1918) 134--139.

\bibitem{Schiff:1968}
L.~I. Schiff, Quantum Mechanics, McGraw Hill, New York, 1968.

\bibitem{Finkelstein:1928}
B.~N. Finkelstein, Über virialsatz in der wellenmechanik, Z.~Phys. 50~(3)
  (1928) 293--294.

\bibitem{Fock:1930}
V.~Fock, Bemerkung zum virialsatz, Z.~Phys. 63~(11) (1930) 855--858.

\bibitem{Albeverio:1972}
S.~Albeverio, On bound states in the continuum of n-body systems and the virial
  theorem, Ann.~Phys. 71~(1) (1972) 167--276.

\bibitem{Weislinger:1974}
E.~Weislinger, G.~Olivier, The classical and quantum mechanical virial theorem,
  Int.~J.~Quantum~Chem. 8~(S8) (1974) 389--401.

\bibitem{Kalf:1976}
H.~Kalf, The virial theorem in relativistic quantum mechanics, J.~Funct.~Anal.
  21~(4) (1976) 389--396.

\bibitem{Gersch:1979}
H.~A. Gersch, Another derivation of the quantum virial theorem, Am.~J.~Phys.
  47~(6) (1979) 555--555.

\bibitem{Leinfelder:1981}
H.~Leinfelder, On the virial theorem in quantum mechanics,
  Integr.~Equ.~Oper.~Theory 4~(2) (1981) 226--244.

\bibitem{Burghardt:2005}
I.~Burghardt, Dynamics of coupled bohmian and phase-space variables: A moment
  approach to mixed quantum-classical dynamics, J.~Chem.~Phys. 122~(9) (2005)
  094103.

\bibitem{Maslov:1972}
V.~Maslov, V.~Arnol'd, V.~Buslaev, Th{\'e}orie des perturbations et
  m{\'e}thodes asymptotiques, Dunod Gauthier--Villars, Paris, 1972.

\bibitem{Arnold:1967}
V.~I. Arnold, Characteristic class entering in quantization conditions,
  Funct.~Anal.~Its~Appl. 1~(1) (1967) 1--13.

\bibitem{Cappell:1994}
S.~E. Cappell, R.~Lee, E.~Y. Miller, On the maslov index,
  Commun.~Pure~Appl.~Math. 47~(2) (1994) 121--186.

\bibitem{deGosson:1997}
M.~A. de~Gosson, Maslov Classes, Metaplectic Representation and Lagrangian
  Quantization, Wiley, Berlin, 1997.

\bibitem{Mcduff:1998}
D.~McDuff, D.~Salamon, Introduction to Symplectic Topology, Oxford University
  Press, Oxford, 1998.

\bibitem{Mahalanobis:1936}
P.~C. Mahalanobis, On the generalised distance in statistics, in: Proceedings
  of the National Institute of Science of India, Vol.~12, 1936, pp. 49--55.

\bibitem{Bertrand:1987}
J.~Bertrand, P.~Bertrand, A tomographic approach to wigner's function,
  Found.~Phys. 17~(4) (1987) 397--405.

\bibitem{Smithey:1993}
D.~T. Smithey, M.~Beck, M.~G. Raymer, A.~Faridani, Measurement of the wigner
  distribution and the density matrix of a light mode using optical homodyne
  tomography: Application to squeezed states and the vacuum, Phys.~Rev.~Lett.
  70 (1993) 1244--1247.

\bibitem{Dariano:1995}
G.~M. D'Ariano, U.~Leonhardt, H.~Paul, Homodyne detection of the density matrix
  of the radiation field, Phys.~Rev.~A 52 (1995) R1801--R1804.

\bibitem{Mancini:1997}
S.~Mancini, V.~I. Man'ko, P.~Tombest, Classical-like description of quantum
  dynamics by means of symplectic tomography, Found.~Phys. 27~(6) (1997)
  801--824.

\bibitem{Manko:2000}
M.~A. Man'ko, Quasidistributions, tomography, and fractional fourier transform
  in signal analysis, J.~Russ.~Laser~Res. 21~(5) (2000) 411--437.

\bibitem{Healy:2015}
J.~Healy, M.~Kutay, H.~Ozaktas, J.~Sheridan, Linear Canonical Transforms:
  Theory and Applications, Springer Series in Optical Sciences, Springer, New
  York, 2015.

\bibitem{Manko:2019}
V.~Man'ko, L.~Markovich, Quantum tomography of time-dependent nonlinear
  hamiltonian systems, Rep.~Math.~Phys. 83~(1) (2019) 87--106.

\bibitem{Zhang:2014}
K.~Zhang, F.~Bariani, P.~Meystre, Theory of an optomechanical quantum heat
  engine, Phys. Rev. A 90 (2014) 023819.

\bibitem{Qian:2012}
J.~Qian, A.~A. Clerk, K.~Hammerer, F.~Marquardt, Quantum signatures of the
  optomechanical instability, Phys. Rev. Lett. 109 (2012) 253601.

\bibitem{Lorch:2014}
N.~L\"orch, J.~Qian, A.~Clerk, F.~Marquardt, K.~Hammerer, Laser theory for
  optomechanics: Limit cycles in the quantum regime, Phys.~Rev.~X 4 (2014)
  011015.

\bibitem{Elouard:2015}
C.~Elouard, M.~Richard, A.~Auff{\`{e}}ves, Reversible work extraction in a
  hybrid opto-mechanical system, New~J.~Phys. 17~(5) (2015) 055018.

\bibitem{Abari:2019}
N.~E. Abari, G.~V.~D. Angelis, S.~Zippilli, D.~Vitali, An optomechanical heat
  engine with feedback-controlled in-loop light, New~J.~Phys. 21~(9) (2019)
  093051.

\bibitem{Bennett:2020}
J.~S. Bennett, L.~S. Madsen, H.~Rubinsztein-Dunlop, W.~P. Bowen, A quantum heat
  machine from fast optomechanics, New~J.~Phys. 22~(10) (2020) 103028.

\bibitem{Malkin:1969}
I.~Malkin, V.~Man'ko, D.~Trifonov, Invariants and the evolution of coherent
  states for a charged particle in a time-dependent magnetic field,
  Phys.~Lett.~A 30~(7) (1969) 414.

\bibitem{Malkin:1973}
I.~A. Malkin, V.~I. Man'ko, D.~A. Trifonov, Linear adiabatic invariants and
  coherent states, J.~Math.~Phys. 14~(5) (1973) 576--582.

\bibitem{Dodonov:1975}
V.~V. Dodonov, I.~A. Malkin, V.~I. Man'ko, Integrals of the motion, green
  functions, and coherent states of dynamical systems, Int.~J.~Theor.~Phys.
  14~(1) (1975) 37--54.

\bibitem{Dodonov:2003}
V.~Dodonov, V.~Man'ko, Theory of Nonclassical States of Light, Taylor $\&$
  Francis, London, 2003.

\bibitem{Dhara:1984}
A.~K. Dhara, S.~V. Lawande, Time-dependent invariants and the feynman
  propagator, Phys.~Rev.~A 30 (1984) 560--567.

\bibitem{Bertin:2012}
M.~C. Bertin, B.~M. Pimentel, J.~A. Ramirez, Construction of time-dependent
  dynamical invariants: A new approach, J.~Math.~Phys. 53~(4) (2012) 042104.

\bibitem{Abdalla:2005}
M.~S. Abdalla, P.~G.~L. Leach, Wigner functions for time-dependent coupled
  linear oscillators via linear and quadratic invariant processes, J.~Phys.~A
  38~(4) (2005) 881--893.

\end{thebibliography}





\end{document}